\documentclass[iop]{emulateapj}

\usepackage{graphicx}
\usepackage{amsmath}
\usepackage{gensymb}
\usepackage{aas_macros}
\usepackage{url}
\usepackage{color}
\usepackage{nicefrac}
\usepackage{float}
\usepackage[caption = false]{subfig}

\newcommand{\red}[1] {#1}

\newcommand{\defeq}{\mathrel{\mathop:}=}

\shorttitle{Ionospheric impact on detection of global EoR}
\shortauthors{Sokolowski et al.}

\begin{document}%

\title{The impact of the ionosphere on ground-based detection of the global Epoch of Reionisation signal}

\author{Marcin Sokolowski$^{1,2}$, Randall B. Wayth$^{1,2}$, Steven E. Tremblay$^{1,2}$, Steven J. Tingay$^{1,2}$, Mark Waterson$^{1,3}$, \\
Jonathan Tickner$^{1}$, David Emrich $^{1}$, Franz Schlagenhaufer$^1$, David Kenney $^{1}$, Shantanu Padhi $^{1}$},
\affil{
$^1$International Centre for Radio Astronomy Research, Curtin University, GPO Box U1987, Perth, WA 6845, Australia; marcin.sokolowski@curtin.edu.au \\
$^2$ARC Centre of Excellence for All-sky Astrophysics (CAASTRO) \\
$^3$SKA Organisation, Jodrell Bank Observatory, Lower Withington, Macclesfield, SK11 9DL, United Kingdom 
}

\begin{abstract}
The redshifted 21cm line of neutral hydrogen (\textsc{Hi}), potentially observable at low radio frequencies ($\sim50-200$\,MHz), is a promising probe of the physical conditions of the inter-galactic medium during Cosmic Dawn and the Epoch of Reionisation (EoR).
The sky-averaged \textsc{Hi} signal is expected to be extremely weak ($\sim$100\,mK) in comparison to the Galactic foreground emission ($\sim 10^4$\,K). Moreover, the sky-averaged spectra measured by ground-based instruments are affected by chromatic propagation effects ($\sim$tens of kelvin) originating in the ionosphere.
We analyse data collected with the upgraded BIGHORNS system deployed at the Murchison Radio-astronomy Observatory to assess the significance of ionospheric effects on the detection of the global EoR signal.
The ionospheric effects identified in these data are, particularly during nighttime, dominated by absorption and emission. We measure some properties of the ionosphere, such as the electron temperature ($T_e \approx$470\,K at nighttime), magnitude, and variability of \red{optical depth} ($\tau_{100 MHz} \approx$0.01 and $\delta \tau \approx$0.005 at nighttime).
According to the results of a statistical test applied on a large data sample, very long integrations ($\sim$100\,h collected over approximately two months) lead to increased signal to noise even in the presence of ionospheric variability.
This is further supported by the structure of the power spectrum of the sky temperature fluctuations, which has flicker noise characteristics at frequencies $\gtrsim 10^{-5}$\,Hz, but becomes flat below $\approx 10^{-5}$\,Hz.
Hence, we conclude that the \red{stochastic error} introduced by the chromatic ionospheric effects tends to zero in an average. 
\red{Therefore, the ionospheric effects and fluctuations are not fundamental impediments preventing ground-based instruments from integrating down to the precision required by global EoR experiments, provided that the ionospheric contribution is properly accounted for in the data analysis.}
\end{abstract}

%
\maketitle%
\section{INTRODUCTION}
\label{sec_intro}

Detection of the highly redshifted 21 cm signal due to the spin-flip transition in neutral hydrogen can be a powerful probe of the early Universe ($z>15$) and the Epoch of Reionisation (EoR; $15 > z > 6$) in particular.
Observational efforts are being undertaken on several fronts ranging from large interferometric arrays to single antenna experiments. 
The latter aim to detect the signature of the EoR in the integrated spectrum of the entire visible sky at low frequencies -- the ``global'' EoR signal.
This measurement is extremely challenging, primarily because the expected signal ($\sim 100$\,mK) is $4-5$ orders of magnitude weaker than the foregrounds ($\sim$1000\,K). The global EoR approach has already resulted in a lower limit ($\Delta z > 0.06$) on the duration of the EoR \citep{2010Natur.468..796B} obtained by the Experiment to Detect the Global EoR Step (EDGES).

Achieving precision on the order of a few milliKelvin requires an extremely stable radiometer system and a nearly perfect understanding and calibration of the signal.
For a given frequency, the required integration time depends on the receiver noise temperature and frequency bin width selected to optimize data analysis (limited by the resolution of a particular spectrometer).
\red{In principle, according to estimates for the Broadband Instrument for Global Hydrogen Reionisation Signal (BIGHORNS; \nocite{bighorns_paper}Sokolowski et al. 2015, hereafter referred to as S15) system with a receiver noise temperature $\sim$90\,K, a statistical error of $\sim$10\,mK in a 1\,MHz bin at frequency 80\,MHz can be achieved in $\sim$4\,hours of integration ($\sim$1mK requires $\sim$400\,hours) at optimal, ``cold'' sky, conditions (i.e. nighttime when the Galactic Center is below the horizon and sky noise temperature at 80\,MHz is $\approx$1200\,K ).}
In practice however, due to the uncertainty of the underlying sky spectrum, which needs to be fitted and subtracted, integration times longer by as much as an order of magnitude are required to achieve residuals of $\sim$10\,mK.
Moreover, precision of $\sim$10\,mK might not be sufficient for data analysis methods to constrain reionization models and recover physical parameters (represented by predicted turning points in the sky-averaged signal) of the transition from the Dark Ages to the reionized Universe \citep{2012AdSpR..49..433B,2012MNRAS.419.1070H}. Therefore, integrating down to $\sim$mK precisions requires averages over data collected during multiple nights. 

This basic estimate, however, does not account for additional \red{stochastic error}\footnote{\red{In order to reflect the nature of ionospheric fluctuations, we will further call the additional error which they introduce ``stochastic error''.}} which may come from instrumental or other effects such as propagation of the sky signal through the Earth's atmosphere. The atmosphere can affect the cosmic signal propagation via several effects. The most important for the global EoR signal experiment are chromatic effects in the ionosphere.
Achromatic refraction in the troposphere is less important because it does not affect the frequency structure of the signal. Moreover, its variability (due to variations in ambient temperature and atmospheric pressure), which we estimate to be $\lesssim$2\,K above 50\,MHz, is at least one order of magnitude smaller than variability of the chromatic ionospheric effects.
Nevertheless, the tropospheric refraction may have an indirect impact through duct formation \citep{tropo_ducting} which enables long distance radio-frequency interference (RFI) propagation and therefore occasionally causes significant data excision rates.
In the following subsections we will present some properties of the ionosphere most relevant to the global EoR experiments.

\subsection{Ionosphere}
\label{subsec_ionosphere}

The ionosphere is the part of the Earth's atmosphere where there is sufficient ionization to affect the propagation of radio waves. 
It consists of several layers: D-layer at height $50-90$\,km; E-layer at $90-140$\,km; and F-layer at heights above 140\,km \citep{Ionospheric_Radio,evans_and_hagfors}.
The most relevant to single antenna, high precision measurements of sky brightness temperature at low frequencies are the D-layer and F-layer.

(i) \textbf{F-layer}: most of the ionospheric electron column density is accounted for by the F-layer, where electron density is \red{$\sim10^{12}$\,$e^{-}/m^{3}$} during the daytime and typically an order of magnitude less during the nighttime \citep{Ionospheric_Radio,evans_and_hagfors}.
Measurements of the total electron content (TEC)\footnote{TEC is the total number of electrons per $m^2$ (electron column density) along a given line-of-sight (typically determined for vertical direction). 1 TEC unit is a column density of $10^{16}$\,$e^{-}/m^{2}$}, routinely monitored by measuring the propagation delay of GPS signals, are dominated by electrons in the F-layer.

\red{Due to very high electron content the F-layer dominates refractive effects, which may play a non-negligible role for global EoR experiments by shifting radio sources within the antenna beam.}
\red{Based on equation 6 in \citet{bailey_1948}, we estimated the average position offsets of radio sources for typical nighttime electron density $10^{11}\,e^{-}/m^{3}$ to be $\sim0.1$\,arcmin at 100\,MHz and zenith angle $\theta=45\degree$ (corresponding to $\sim0.3$\,arcmin at 100\,MHz and zenith angle 60$\degree$ cited by \citet{THOMPSON}).}
\red{According to the Bailey's equation, the expected intra-day variations of the average offset are proportional to variations in electron density of typically $\sim$20-30\% (see Section~\ref{subsec_contrib_iono_refr}).}
\red{Using the data from the GLEAM survey \citep{gleam} performed with the MWA telescope \citep{mwa} located at the Murchison Radio-astronomy Observatory (MRO), \citet{balwinder} report nighttime offsets of the source positions of the order of a fraction of arcmin at 100\,MHz and scaling as $1/\nu^2$.}
\red{On the other hand, \citet{cleo_loi} report slightly higher $\sim$1\,arcmin source offsets at 100\,MHz due to electron density gradients (plasma tubes) in the ionosphere, but such prominent effects were only observed occasionally and therefore can be treated as a non-typical behavior.}

On the other hand, signal loss in the ionosphere measured in dB is expected to be of the order of $L_{dB} \sim 10^{-22}\bar{\nu_c} TEC_L (100/\nu_{MHz})^2$ \citep{evans_and_hagfors}, where $\bar{\nu_c}$ is the mean rate of electron collisions, and $TEC_L$ is the mean total electron content in the layer L, expressed as a product of mean electron density ($N_e$) and the width of the ionospheric layer ($\Delta h_L$).
Therefore, due to low collision rates $\nu_c \sim 10^2$\,Hz \citep{Thrane1966721,Aggarwal1979753}, nighttime absorption in the F-layer is considered to be smaller than in the D-layer; but due to high electron density it can be significant during the daytime \citep{steiger_and_walter}.

(ii) \textbf{D-layer}: is the lowest part of the ionosphere where electron densities are $\sim$2-3 orders of magnitude lower than in the F-layer. High electron densities in the D-layer exist only during the daytime due to insolation. 
Typical daytime electron density in the D-layer is $\sim10^9$\,$e^{-}/m^{3}$, falling to a residual density of $\sim 10^8$\,$e^{-}/m^{3}$ at night. Due to significant collision rates of the order of $\nu_c \sim 10^6$\,Hz \citep{collision_freq_in_D_region,Aggarwal1979753,Ionospheric_Radio} the D-layer dominates nighttime ionospheric absorption.
The nighttime residual electron density causes absorption effects of order $L_{dB}\sim L_{dB}^{100 MHz} \times (100/\nu_{MHz})^2 \sim 0.01$\,dB$\times (100/\nu_{MHz})^2$ at zenith \citep{evans_and_hagfors,Ionospheric_Radio,THOMPSON}, which may have a significant impact on global EoR measurements.
\red{The magnitude of typical nighttime absorption at 100\,MHz, $L_{dB}^{100 MHz} \sim 0.01$\,dB, corresponds to optical depth of the ionosphere $\tau \approx 0.0023$.}

(iii) \textbf{Thermal emission}:  the ionosphere emits thermal radio emission \citep{Hsieh1966783} with brightness temperature $T_{iono}\sim \tau T_{e}$, where $\tau$ is the \red{optical depth} of the ionosphere and $T_{e}$ is the electron temperature.
Because $T_{iono}$ is proportional to \red{optical depth} it is expected that nighttime emission is dominated by the D-layer, whilst during the daytime there should be a significant contribution from the F-layer.
The electron temperature in the ionosphere has been measured with a variety of methods ranging from rocket measurements \citep{Oyama20001231} to ground-based observations.
The electron temperatures of the lower ionosphere (D-layer) measured at frequencies 1.6 and 2\,MHz, at which radio signals from outer space are entirely reflected, are typically in the range $\sim240 - 290$\,K \citep{Pawsey}.
The electron temperatures at higher altitudes (F-layer), measured by \citet{zhang_holt_zalucha} using incoherent scatter radar (ISR) experiments, are typically $\sim 800-1000$\,K at $125\le h \le 175$\,km and $\sim 1000-2000$\,K at $275\le h \le 325$\,km, and exhibit seasonal and solar cycle variations.

Due to these effects the ionosphere, besides affecting ground based and satellite radio-communications, significantly influences low frequency radio signals from outer space observed on the ground.
Moreover, the properties of the ionosphere vary in time on multiple timescales due to factors affecting the amount of ionization such as, for example, the solar cycle\footnote{Beginning of 2015 is just after the maximum of cycle 24 (http://solarscience.msfc.nasa.gov/predict.shtml)}, solar activity on a particular day, or cosmic-ray flux.
All these effects have resulted in an on-going discussion in the literature on the importance of ionospheric effects \citep{harish_et_al} and whether the global EoR signature can feasibly be detected from the ground \citep{datta_et_al}.

\subsection{Ionospheric impact on the global EoR detection}
\label{subsec_ionosphere_impact}

The impact of the aforementioned ionospheric effects on the detection of the global EoR has recently been investigated by \citet{harish_et_al} and \citet{datta_et_al}.
These studies have shown that ionospheric absorption and refraction can significantly affect the frequency structure of the cosmic signal.
\citet{datta_et_al} conclude that long timescale variability of ionospheric absorption and refraction resulting from observed TEC variability may impede even perfect ground-based instruments from integrating down to the required $\sim$mK precision.
Hence, it is vital to identify ionospheric effects in the sky-averaged radio spectrum at the frequencies of interest ($50 - 200$\,MHz) and assess their significance for global EoR experiments.
The observation of ionospheric effects at frequencies $80-185$\,MHz in the sky-averaged radio spectrum have recently been reported by the EDGES group \citep{2014arXiv1412.2255R}.
The EDGES experiment measured variability of absorption (\red{optical depth}) and magnitude of emission (electron temperature) of the ionosphere, which proves that these effects do play an important role and must be accounted for in data analysis. 

Additionally, such measurements present an interesting alternative technique to measure the properties of the ionosphere at low radio frequencies with wide-band total power radiometers.
Typically, relative absorption is measured by narrow-band relative ionospheric opacity meters (riometers)\footnote{\red{Throughout this manuscript we use the term ``optical depth'' for the measured absorption. However, we find the naming in the riometer field confusing as people typically talk about ``opacity'' and even the term riometer is an acronym for a relative ionospheric ``opacity'' meter, whilst in fact everybody seems to measure what is strictly speaking ``optical depth''. Probably, like in many other fields this is some historical relic, but we did not get to the bottom of the problem.}} at lower frequencies of $\sim 30$\,MHz \citep{riometer_little} or 18.3\,MHz \citep{Mitra_et_al}. 
\red{Based on multi-day observations, a so-called `quiet day curve' (QDC) of the maximum power observed by the riometer during a state of very small, residual ionospheric absorption (typically during the pre-dawn hours \citep{belokovich}) is determined and relative absorption is measured with respect to the QDC \citep{qdc_method}. }
Similar techniques can be applied to wide-band sky-averaged radio spectra from global EoR experiments, which makes them useful tools for ionospheric studies.
Moreover, wide-band data from these instruments enables analyses of the frequency structure of ionospheric absorption and emission, which are not available from traditional narrow band riometers. 
Finally, applications of the riometer approach are not limited to ionospheric studies, as they turned out to be useful tools even for inferring X-ray flux of the largest solar X-ray flare ever recorded \citep{brodrick}.

This paper presents an analysis aiming to assess the significance of ionospheric effects on the detection of the global EoR signal.
Data were collected in 2014-2015 by the BIGHORNS system at the Murchison Radio-astronomy Observatory (MRO) in Western Australia.
In Section~\ref{sec_instrument} we describe the BIGHORNS system as it was deployed at the MRO in October 2014, mainly focusing on modifications with respect to the system presented in S15. 
In Section~\ref{sec_data_analysis} we present data processing and data analysis leading to identification of the ionospheric effects in the data. We compare these results to the expected contributions from ionospheric absorption/emission and refraction.
We also present some properties of the ionosphere, such as electron temperature, relative \red{optical depth}, and its variability derived from the data analysis.
In Section~\ref{sec_data_modelling} we compare our derived electron temperature with model predictions.
Finally, in Section~\ref{sec_stddev_test} we perform a statistical test to verify whether the observed data can be integrated down to the required $\sim$\,mK precision by averaging large numbers of spectra in the presence of a variable ionosphere.
We compare observed variability in the real data with the variability expected for diurnal changes in antenna temperature due to Earth rotation. In order to better understand the results of this test in the context of the results of \citet{datta_et_al}, we also study the power spectrum of observed fluctuations in the sky temperature.

\section{INSTRUMENT DESCRIPTION}
\label{sec_instrument}

\begin{figure}
  \begin{center}
    \includegraphics[width=3in]{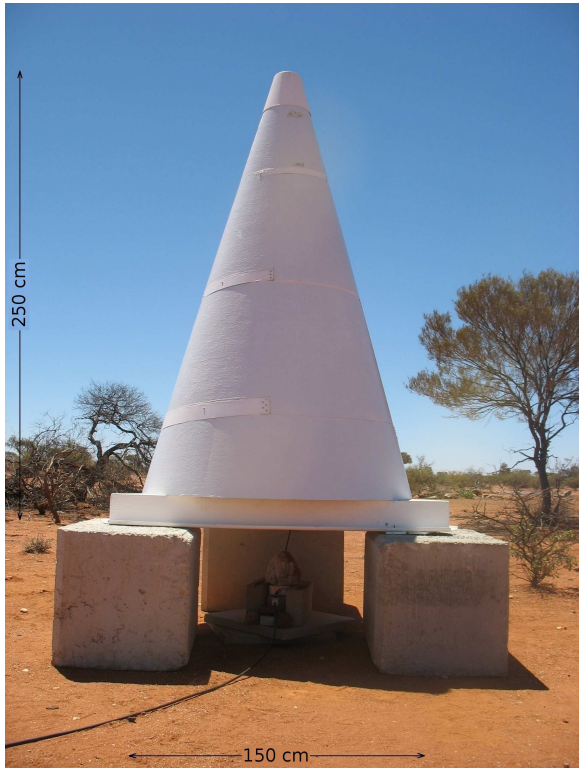}
    \caption{The conical log-spiral antenna as deployed at the MRO in October 2014. The antenna was painted in order to protect the copper conductor from the weather.}
    \label{fig_cone_at_the_mro}
  \end{center}
\end{figure}

The detailed description of the initial installation of the BIGHORNS total power radiometer can be found in S15. The BIGHORNS system in a very similar form was deployed at the MRO in October 2014.
The major difference with respect to the system described in S15 is the antenna. The portable biconical antenna, which was used during short remote deployments, was replaced with a conical log-spiral (CLS) antenna (Fig.~\ref{fig_cone_at_the_mro}). 
The antenna was painted white in order to protect the copper conductor from oxidizing and to protect it from bad weather conditions. The balun is installed at the apex of the cone under a protective cap and is connected to the front-end receiver via a cable of approximately 2.5\,m length.

\begin{figure}
  \begin{center}
    \includegraphics[width=3in]{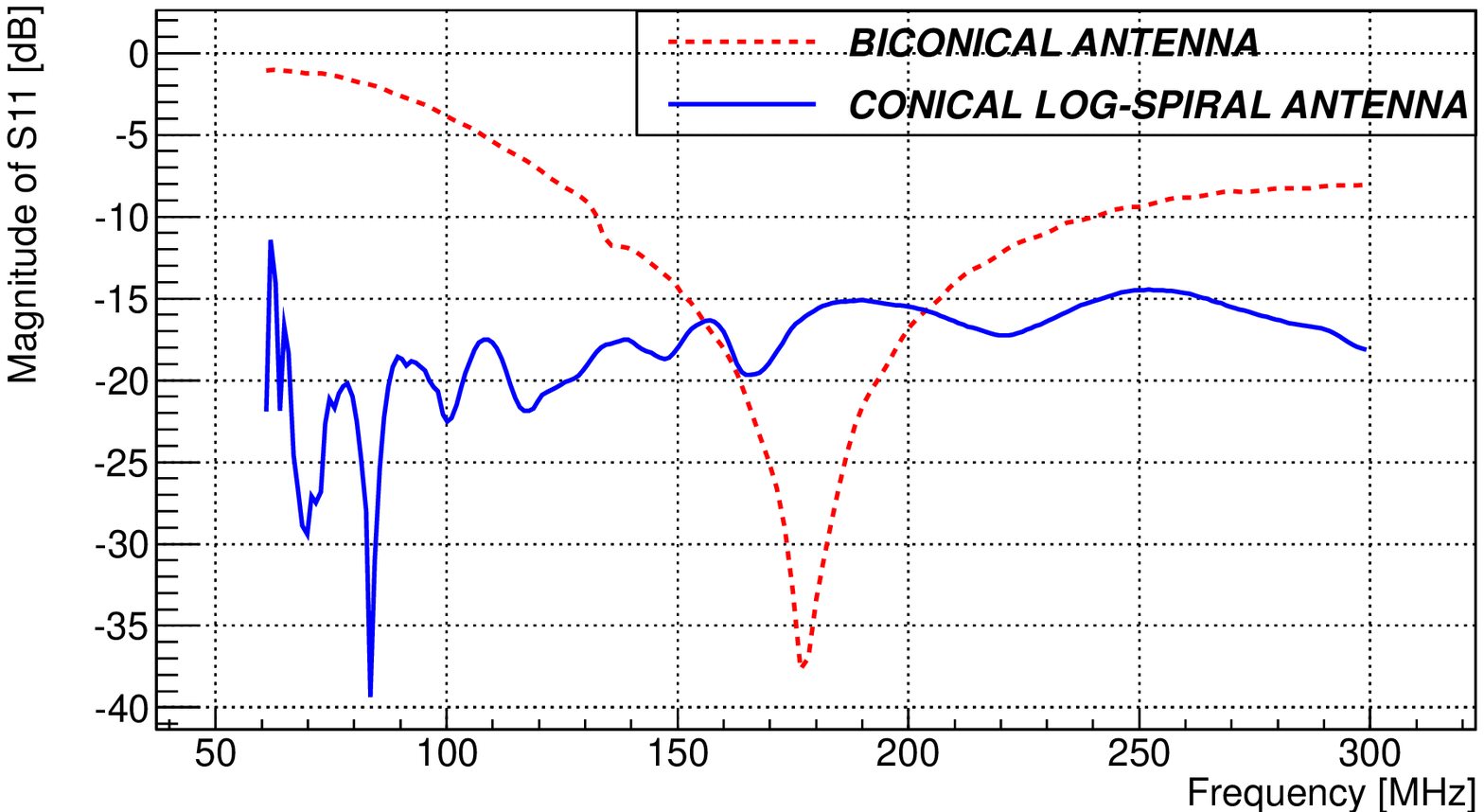}
    \includegraphics[width=3in]{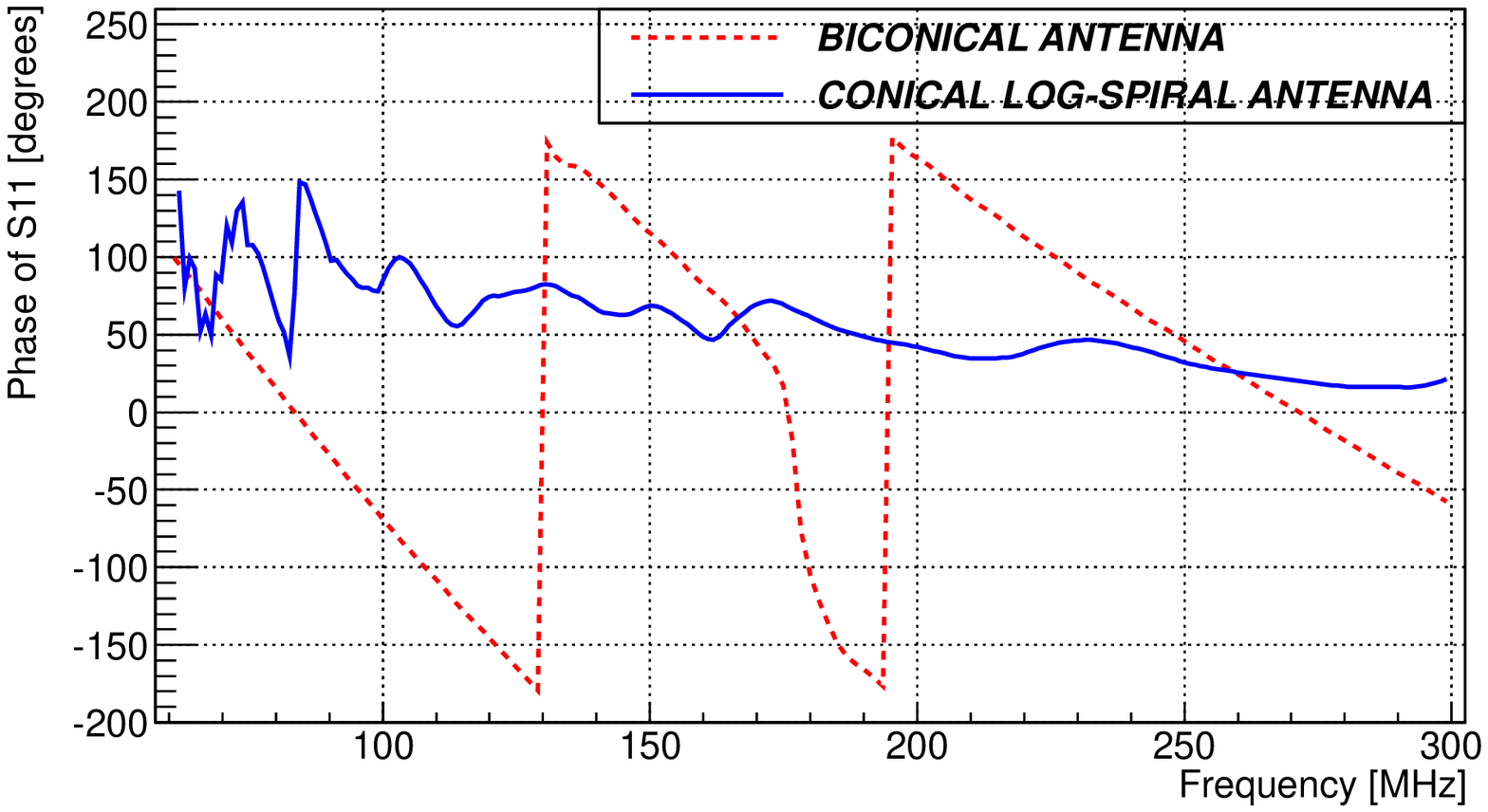}
    \caption{\red{Comparison of the reflection coefficient magnitudes (upper figure) and phases (lower figure) of the log spiral antenna and previously used biconical antenna.}}
    \label{fig_s11mag_cone_vs_bicon}
  \end{center}
\end{figure}

\red{The design details of the antenna and motivation for the upgrade can be found in \citet{aziz_thesis} and S15. In summary, the upgrade was motived by the following reasons.}
\red{Firstly, the response (beam pattern and the reflection coefficient) of the CLS is, by design, more frequency independent than the response of the previously used biconical antenna.}
\red{Secondly, it is very well matched (\mbox{$\le$-15\,dB}) across a wide frequency band ($\sim$70-180\,MHz), which improves signal to noise ratio and facilitates a more accurate calibration over a wider frequency band. Particularly, the CLS antenna significantly improves sensitivity at low frequencies (70\,MHz\,$\lesssim \nu \lesssim$\,150\,MHz), which are of main interest for the EoR experiment. Nevertheless, its reflection coefficient has more frequency structure (i.e. ripples or spikes) compared to the biconical antenna (Fig.~\ref{fig_s11mag_cone_vs_bicon}). This frequency structure, which results from the constructive and destructive interference of the RF-signals reflected from the unterminated ends of the spiral arms, may pose a problem for the data analysis looking for a tiny ($\sim$\,mK) global EoR ``ripple'' amongst the foregrounds (assumed to be smooth as a function of frequency). The comparison of the reflection coefficients of the biconical and CLS antennas, which were measured at the end of the cables connecting the balun to the front-end receiver, is presented in Figure~\ref{fig_s11mag_cone_vs_bicon}.}
\red{Thirdly, the exact knowledge and accessibility of all its components (such as the balun, output cable etc.) enabled us to better characterize the CLS antenna.}
\red{Finally, the CLS antenna is a left-hand circularly polarized (LHCP) antenna which slightly helped in suppressing reception of undesired right-hand circularly polarized (RHCP) signals from the ORBCOMM satellites. We estimate that the CLS antenna attenuates the undesired ORBCOMM signals (at 137-138\,MHz) by a few dB (see also Section~\ref{subsec_initial_data_processing}).}

\begin{figure}  
  \begin{center}
    \includegraphics[width=3in]{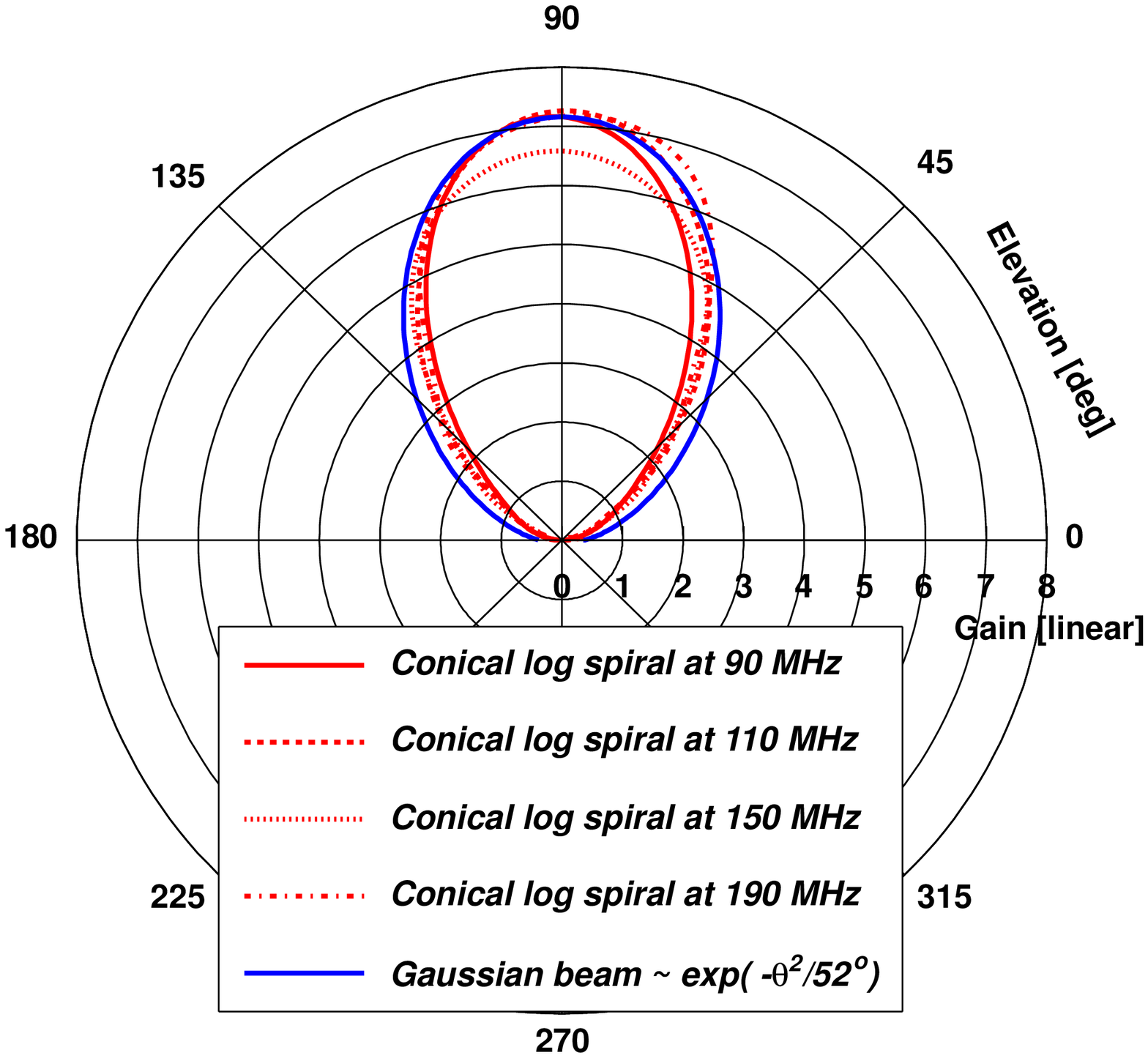}
	 \includegraphics[width=3in]{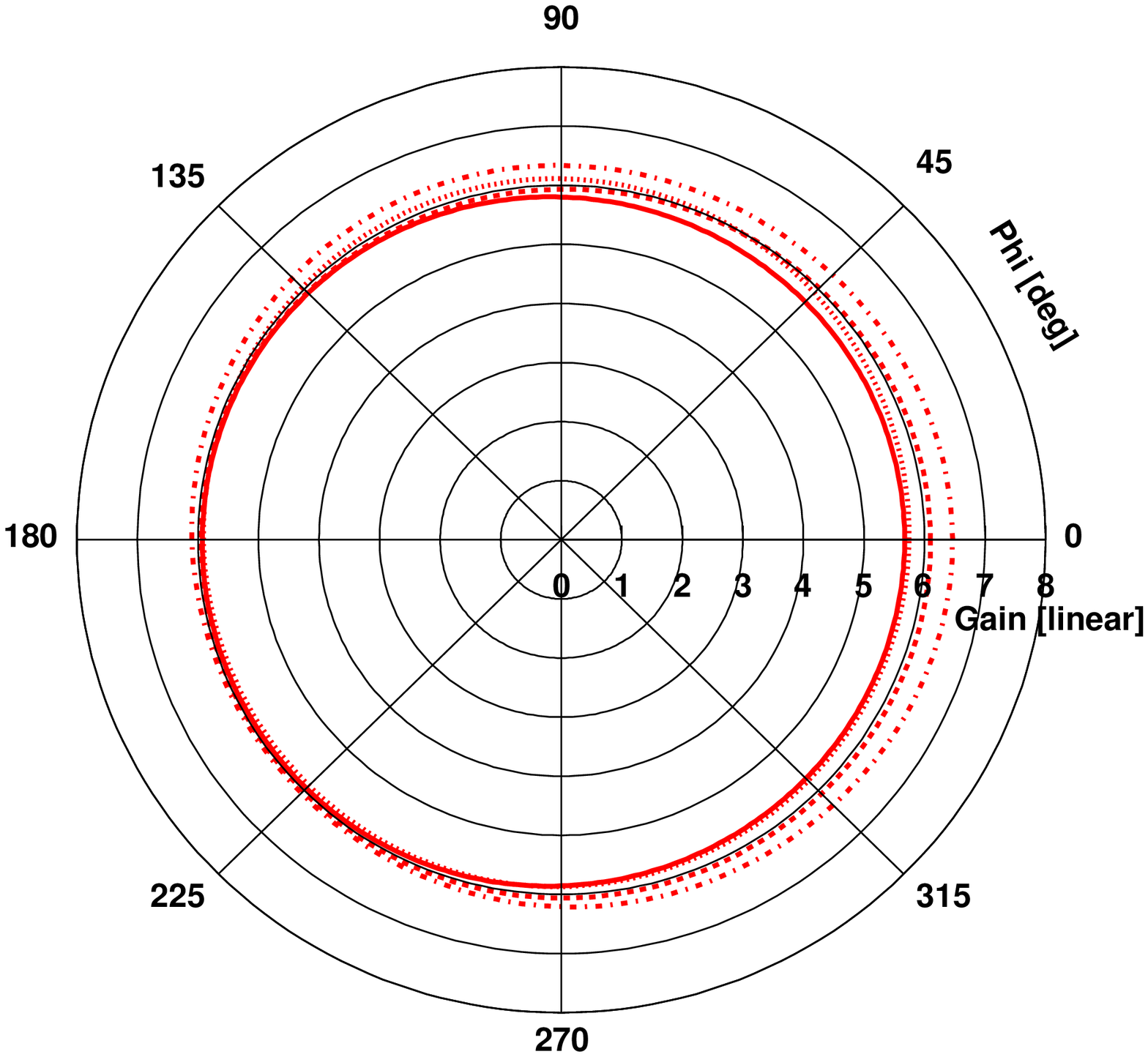}
    \caption{\red{The antenna pattern of the conical log spiral antenna simulated in the FEKO software. The upper figure shows the simulated CLS antenna gain as a function of elevation angle at at 90, 110, 150, and 190\,MHz (red lines) with a Gaussian beam gain$\sim$\,$exp(- (\theta/52\degree)^2 )$ (blue line). The lower figure shows antenna gain as a function of azimuthal angle at elevation angle 70$\degree$. The antenna response is within $\sim10\,\%$ frequency and azimuthal angle independent.}}
    \label{fig_cone_pattern_feko}
  \end{center}
\end{figure}

\red{The antenna was modeled in the FEKO simulation package. Its antenna pattern is within $\sim$10\% frequency independent and symmetrical in azimuthal angle ($\phi$) in 75-200\,MHz band (Fig.~\ref{fig_cone_pattern_feko}). Hence, we will further approximate its response as frequency and azimuthal angle independent.}
Upgrading to the better-matched antenna allowed us to remove the 3\,dB attenuator which was originally installed between the biconical antenna and the front-end box to suppress the effects of the impedance mismatch between the two.
Therefore, the \red{receiver noise temperature of the new system (without any additional attenuation between the antenna and front-end) was reduced to $T_{rcv}\approx$180\,K}.

\section{DATA ANALYSIS}
\label{sec_data_analysis}

The system collected data in exactly the same mode as described in S15.
Therefore, the data acquisition parameters will be only briefly reviewed here.
In the present configuration the system uses cold calibration schema and switches between the antenna (for $\approx15.5$\,sec) and a reference source (for 5\,sec), which is a 50\,$\Omega$ terminator at ambient temperature (measured with a portable temperature logger).
The data were collected continuously with 100\% duty cycle (no gaps in time\footnote{The recorded data are continuous, but the sky data (collected on the antenna) have 5\,sec gaps due to the aforementioned switching to a reference source.}) at approximately 50\,ms time and 117.2\,kHz frequency resolution.
\red{The FPGA based spectrometer measured the spectra in $0-480$\,MHz band and the analog filters limited the bandwidth to $\approx 40-350$\,MHz.} The power spectra produced by the spectrometer were saved to FITS files \citep{2010A&A...524A..42P}, which consisted of interleaved blocks of antenna and reference data.
Each FITS file consisted of 1000 integrations, thus every file covered a 50\,sec interval.

\subsection{Initial data pre-processing and calibration}
\label{subsec_initial_data_processing}

The details of initial data pre-processing can be found in S15. In summary, it consisted of the following main steps: 

\begin{enumerate}   
\item \textbf{State identification} - The calibration RF-switch operates autonomously. Therefore, the first data processing step identifies integrations collected on the antenna and reference source and saves the location of the transitions to FITS file headers.
\item \textbf{RFI excision} - The entire integration was excised if the power measured by the spectrometer exceeded a threshold of -35\,dBm \red{($\approx3 \times 10^{-4}$\,mW)}\red{\footnote{\red{dBm is the ratio of measured power referenced to 1\,mW and expressed in decibels, $P_{dBm} = 10$ $log_{10} (P_{mW})$}}} in any single channel or the sum of power in the entire band was above the median galactic noise level plus the threshold of -37.5\,dBm \red{($\approx1.8 \times 10^{-4}$\,mW)}.
The data accepted by these two criteria were flagged by the AO-flagger software \citep{2012A&A...539A..95O}, in order to identify individual channels affected by RFI. Finally, the data were checked with variability criteria (see Sec.~\ref{subsec_solar_trigger}) in order to exclude significantly variable data (due to solar activity or lightning).
\red{The average excision rate due to high power ($\ge$-35\,dBm) from the ORBCOMM satellites (at 137-138\,MHz) of the system with the CLS antenna (without the 3\,dB attenuator between the antenna and the front-end) was $\approx$0.7\% comparing to $\approx$10.9\% with the previously used biconical antenna (after adding 3\,dB due to an additional attenuator). Therefore, based on shift of the two power distributions, we estimate that the CLS antenna suppresses the undesired ORBCOMM signals by a few dB ($\sim$3-5\,dB).}
\item \textbf{Data reduction} - The data unflagged by the RFI excision procedure were reduced to a more manageable volume by averaging (with a median estimator) every \red{$N_{ant}=700$ integrations on the antenna accepted by the RFI excision criteria} and $N_{ref}$ in the range $200 - 300$ on the reference load.
After this step the data formed an averaged dynamic spectrum where each integration was an average of \red{$\tau_{ant}= 700 \times 0.05 = 35$\,sec} of single integrations in the case of antenna and $\tau_{ref} \approx 12.5$\,sec in the case of reference source.
The dynamic spectra from the antenna and reference were saved into separate FITS files.
\red{To enable correct error propagation, the numbers of averaged individual 50\,ms integrations were saved to FITS files (also separate for the antenna and reference data) for every averaged integration and every frequency channel.} 
\red{Hence, calculation of the statistical error on the averaged spectra used actual integrations times (after excision of RFI affected data).}
The resulting reduced data were split by calendar months and saved into separate FITS files representing dynamic spectra for each calendar month. Further data analysis was performed on the reduced dataset.
\item \textbf{Calibration} - each averaged antenna integration in the dynamic spectrum of a given month was calibrated by the corresponding averaged reference integration, according to the calibration procedure described in S15. \red{The procedure uses the power measured on the reference source and information from the ambient temperature probe to calibrate the antenna signal in units of temperature; next, the reflection coefficients (of the antenna and the front-end receiver) are used in order to correct for the reflection of the sky signal from the input of the receiver; finally, the receiver noise temperature and the input noise of the receiver reflected back and forth between the antenna and the receiver (so called reflected noise) are subtracted.}

The only major difference in the calibration procedure was that the reflection coefficient of the conical log spiral antenna measured at the MRO (Fig.~\ref{fig_s11mag_cone_vs_bicon}) was used instead of characteristics of the biconical antenna and there was no additional attenuation between the antenna and the front-end.
As a consistency check, the calibrated data were compared with a sky model \citep{2008MNRAS.388..247D} integrated with an antenna pattern simulated in the FEKO electromagnetic simulation software.
The relative difference between the predicted antenna temperature from the model and the calibrated data is approximately $10\%$, which is consistent with uncertainties in the model temperature scale.
\red{The resulting calibrated dynamic spectra and corresponding statistical errors (calculated according to equation 13 in S15 for appropriate integration times on the antenna and reference) were saved to separate FITS files.
The contribution of the statistical error of averaged reference integrations ($\tau_{ref} \approx 12.5$) to the overall budget of the statistical error is $\approx$50\%, and may be improved in the future in order to reduce the required integration time.}
Therefore, after calibration, two fits files containing the dynamic spectrum and its error were created for each analyzed month (October/November 2014\footnote{In October 2014 only a few days worth of data were collected. Therefore, these data were analyzed together with November 2014 data (forming a 38\,day dataset).}, December 2014, and January 2015).
\end{enumerate}

Further data analysis presented here was performed separately on each calibrated monthly dynamic spectra. 

\subsection{Excision of significantly variable data}
\label{subsec_solar_trigger}

The data processing described in S15 solely focused on nighttime data without any consideration of daytime data, which is regarded as low value for the global EoR analysis.
However, in the presented analysis we aim to derive some properties of the ionosphere as a function of local time including the daytime.
Therefore, in order to include only quiet-Sun data and exclude data affected by solar activity we needed an effective way of excluding data affected by variations on timescales of seconds and less.
Some part of these data are rejected by the AO-flagger software in the RFI excision process, but in order to further clean the data sample we developed a procedure which identifies and flags data affected by short timescale variability.
The procedure acts on individual FITS files (1000 integration each) at 50\,ms resolution after the state identification procedure and is based on the radiometer equation:
\begin{equation}
\red{\sigma_{exp}(\nu) = \frac{P(\nu)}{\sqrt{B \tau_{int}}},}
\label{eq_radiometer_equation}
\end{equation}
where $P(\nu)$ is the power as a function of frequency measured by the radiometer system in arbitrary units, $\sigma_{exp}(\nu)$ is the standard deviation of the measured power, $B$ is the resolution of the frequency bin\red{, and $\tau_{int}$ is the integration time.}
In the absence of short timescale variability, such as solar activity or lightning, the expected standard deviation ($\sigma_{exp}$) as a function of frequency can be calculated from the mean or median spectrum according to equation~\ref{eq_radiometer_equation}.
On the other hand, the actual standard deviation ($\sigma_{obs}$) can be calculated from a sample of antenna integrations. 

\begin{figure}
  \begin{center}	
	\includegraphics[width=3.4in]{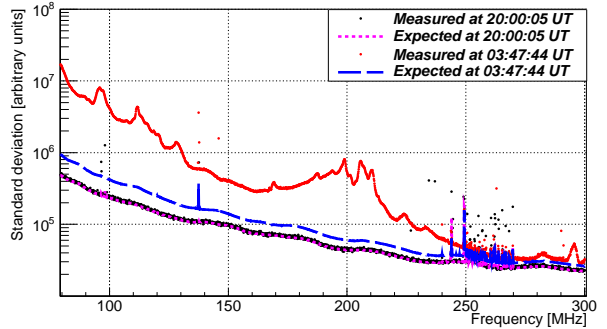}
    \caption{\red{Example of difference between observed (black and red points) and expected (magenta and blue dashed lines) standard deviation of 35\,sec data sample. The black points represent the observed standard deviation of the nighttime data sample collected on 2015-01-16 at 04:00:05 AWST and the magenta dashed line lies on top of these data over almost entire bandwidth (except RFI affected channels above 240\,MHz). The red points, representing the observed standard deviation of the daytime data sample collected on 2015-01-15 at 13:47:44 AWST in the presence of a solar flare, lie much above the dashed blue line corresponding to the standard deviation expected for these data. At that time the Galaxy was transiting during the daytime, therefore the expected standard deviation for daytime (blue long-dashed line) is above the standard deviation expected for the nighttime (magenta short-dashed line). }}
    \label{fig_solar_activity_solarflare_log}
  \end{center}
\end{figure}

\begin{figure}
  \begin{center}
	\includegraphics[width=3.4in]{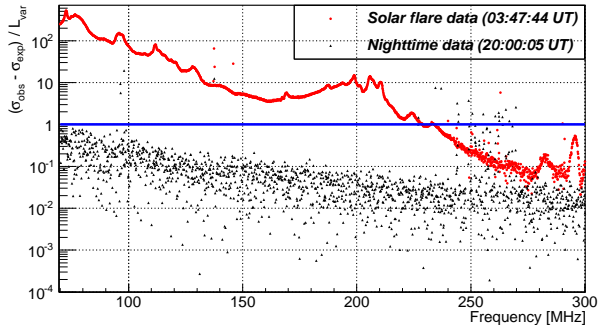}
    \caption{\red{The difference of observed and expected standard deviation of 35\,sec data sample normalized by the threshold ($L_{var}$). The red dots represent the data collected on 2015-01-15 at 13:47:44 AWST in the presence of a solar flare and the black triangles represent the nighttime data collected on 2015-01-16 at 04:00:05 AWST. The data points above the blue horizontal line are considered as affected by variability.}}
    \label{fig_solar_activity_ratio}
  \end{center}
\end{figure}

In the presented analysis we calculated the observed standard deviation from all the antenna integrations in a single FITS file (typically $\sim 700 \times$50\,ms integrations). 
\red{The main goal of this criterion was to identify and excise any kind of short timescale variability ($\lesssim35$\,sec) such as solar activity (second and even millisecond timescales) or lightning (even sub-millisecond timescales). Longer timescale variability ($\gtrsim$35\,s) was most likely not identified by this procedure. Nevertheless, lack of significant variability on longer timescales was verified on the dynamic spectrum normalized by a median 24\,h (such as Fig.~18 in S15). }
\red{The same integrations were used to calculate the mean power spectrum $P(\nu)$ and the expected standard deviation $\sigma_{exp}(\nu)$ was calculated according to equation~\ref{eq_radiometer_equation}.}
\red{Under nighttime conditions and in the absence of any short timescale variability (e.g. nearby lightning) the expected and observed standard deviation curves lie on top of each other (Fig.~\ref{fig_solar_activity_solarflare_log}). }
\red{However, in the presence of solar flares or thunderstorms the observed standard deviation is significantly larger than calculated according to equation~\ref{eq_radiometer_equation} (Fig.~\ref{fig_solar_activity_solarflare_log}). Therefore, the difference between the observed and expected standard deviation is a good metric of any short timescale variability in the data (Fig.~\ref{fig_solar_activity_ratio}).}

\begin{figure}
  \begin{center}
	 \includegraphics[width=3in]{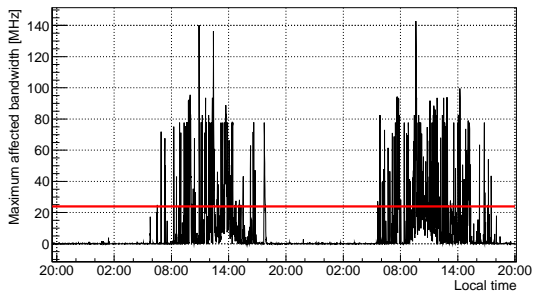}
    \caption{Example of variability metrics (maximum affected bandwidth) as a function of local time for a sample of data collected between 2014-11-01 20:00 and 2014-11-03 20:00 AWST. Clearly the nighttime data are very quiet and daytime data show significant variability mainly due to solar activity. Occasionally nighttime data (not show in this figure) were also significantly affected by nearby lightning. The data exceeding the variability bandwidth threshold of $\approx$24\,MHz (marked in the image with a red line) were excluded from \red{further} analysis.}
    \label{fig_solar_activity_vs_time}
  \end{center}
\end{figure}

\red{The maximum number of consecutive channels $N_{max}$ (corresponding to a maximum affected bandwidth $B_{max}$) in a frequency band $60 - 300$\,MHz where the difference exceeded a frequency independent threshold $L_{var}=5 \times 10^4$ (corresponding to $5\sigma$ around frequency 85\,MHz) was used as a metric to identify broadband short timescale variability.}
\red{The threshold was frequency independent in order to simplify the procedure. However, the method can be further refined and made more sensitive by using a frequency dependent value of the $L_{var}$ threshold.}
Several frequency bands, for example 88-108\,MHz (FM band), 133.9-134.4\,MHz (aeronautical band), 137-138\,MHz (ORBCOMM satellite communication band), 242-272\,MHz (satellites), and multiple other narrow ($\sim$1\,MHz) bands known for RFI were excluded from the procedure.
The maximum frequency bandwidth, $B_{max}$, affected by variability exceeding the \red{$L_{var}$} threshold as a function of local time is shown in Figure~\ref{fig_solar_activity_vs_time}.
In the later data analysis, the data were considered to be affected by short timescale variability if $N_{max}$ for given time exceeded a threshold of 200, which corresponds to a bandwidth of $B_{max} \approx 24$\,MHz.
\red{We have tested several values of the algorithm parameters and found that $N_{max}=200$ ($B_{max} \approx 24$\,MHz) is optimal to suppress false identifications (false positives) due to noise fluctuations in conjunction with RFI and retain a reasonable amount of daytime data (day dependent). The broadband requirement (at least $N_{max}$ channels affected) together with the threshold $L_{var} > 5\sigma$ ensure that the probability of false positives due to noise fluctuations is nearly zero. Nevertheless, in order to improve daytime data analysis, the procedure can be fine-tuned in order to improve its sensitivity to low-level variability.}

\subsection{Data processing}
\label{subsec_data_analysis}

Our analysis of the ionospheric effects in the sky-averaged radio spectrum is based on an approach similar to riometer observations of relative absorption.
In principle, nighttime spectra collected at a given local sidereal time (LST) in the absence of nearby lightning, instrumental, or propagation effects should be the same for each sidereal day.
Therefore, any variation of the same LST sky spectra between days is an indication of variability due to instrumental or local propagation effects.

The problem becomes more complex for daytime data because of the Sun's contribution. Even the quiet Sun introduces small changes to sky-averaged spectra, because its position at a given LST changes slightly every day and differs by $\sim30$\,deg between the first and the last day of the month.
Thus, for a given LST, the solar position within the antenna beam may be significantly different and therefore the solar contribution to the sky-averaged spectrum is also different.
The effect was estimated to be of the order a few Kelvins (depending on Sun's elevation) and was not taken into account in the presented analysis, which may have a small impact on daytime data analysis.
Solar activity events were excised from the analyzed daytime data sample using the methods described in Section~\ref{subsec_solar_trigger}. However, low level solar activity could potentially remain unflagged and contaminate the daytime data sample.
A more detailed subtraction of the solar contribution could further improve analysis of daytime data and is planned in the future.

\begin{figure*}
  \begin{center}
    \includegraphics[width=7in]{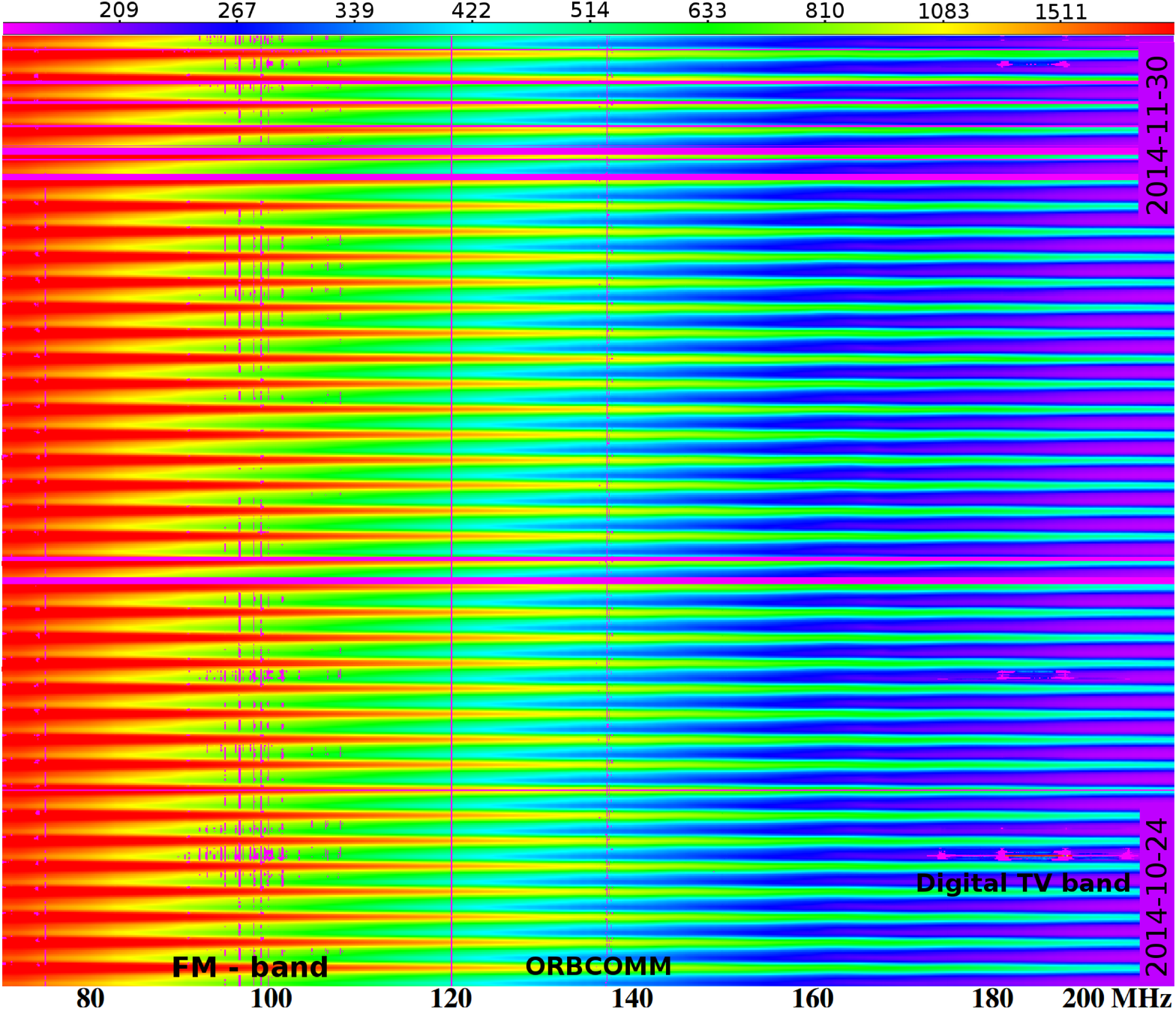}\\ 
    \includegraphics[width=7in,height=3.5in]{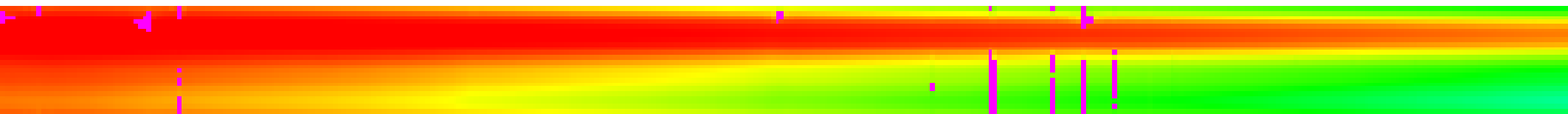}
    \caption{Dynamic spectrum of Oct/Nov 2014 ($\approx$38 days) data in 1\,h LST resolution (upper figure) and a median 24\,h day (lower figure). The median dynamic spectrum is stretched in the time axis. Each 1\,h LST bin results from the median of individual $\approx$35\,sec integrations accepted by RFI and variability criteria, and magenta (zero value) marks data excised by these criteria (for example in the FM-band around 100\,MHz).
The 120\,MHz channel was entirely excised due to self-generated interference in the spectrometer. The horizontal magenta lines correspond to data excised by the variability and RFI criteria. The red ``wedges'' are due to Galaxy transits at LST$\approx$17.8\,h.}
    \label{fig_1h_dynamic_spectrum}
  \end{center}
\end{figure*}




Calibrated dynamic spectra from each calendar month were analyzed separately and integrations accepted by variability criteria (Section~\ref{subsec_solar_trigger}) were averaged in 1\,hour LST bins ($0-1$\,h, $1-2$\,h, $2-3$\,h and so on) forming a dynamic spectrum at 1\,h resolution.
In order to exclude outlier points due to residual RFI, the median estimator was used to average individual $\approx$35\,sec integrations within 1\,h LST bins.
Thus, after this step the dynamic spectrum consisted of $\sim24 \times N_{days}$ individual spectra, where $N_{days}$ is the number of days in the analyzed calendar month (Fig.~\ref{fig_1h_dynamic_spectrum}).

\subsection{Data sample cleaning}  
\label{subsec_data_sample_cleaning}

For each 1\,h LST bin some additional characteristics, such as the mean and standard deviation of the ambient temperature and the maximal variability metric ($N_{max}$) of the individual integrations within the 1\,h LST bin, were also calculated.
Based on this information, the following quality criteria were imposed in order to select only a consistent sample of 1\,h LST bins not affected by instrumental or environmental effects:

\begin{enumerate}
\item \textbf{Ambient temperature}\\
The FEKO software simulations indicated that diurnal variations in ambient temperature may affect the instrumental response - mainly antenna efficiency.
In the analyzed period, Oct 2014 - Jan 2015, the ambient temperature measured at the MRO varied in the range 285\,K - 320\,K. Therefore in order to suppress temperature-related instrumental effects to a minimum, only 1\,h LST bins collected at ambient temperature $T_{amb}$ within $T_{best} \pm 5$\,K were selected for further analysis, where $T_{best}$ is an ambient temperature which maximizes the number of spectra in the sample (different for every LST time). Typically a few ($\le$7) 1\,h LST bins were excluded for every analyzed month.

\item \textbf{Minimum integration time \red{per LST bin}}\\
For each 1\,h LST bin \red{in a day's data}, the total integration time on the antenna $\tau_{1h}$ was calculated as $\tau_{1h} = N \times 35$\,sec, where N is the number of 35\,sec integrations.
It is expected to be less than 1\,h because $\approx$26\% of time was spent on a reference source and some fraction of the data were excised due to solar activity or RFI.
Based on the distribution of $\tau_{1h}$ for all 1\,h LST bins, we determined that $\tau_{1h}=2500$\,s with a standard deviation $\approx$100\,s.

The sky temperature changes significantly within a 1\,h interval (as the Galaxy moves across the sky). Thus, large gaps in the data were undesired because they caused the averaged spectrum of 1\,h to be significantly different than expected for a fully sampled 1\,h period.
For example, assume the Galactic Center was rising within a given 1h LST bin and therefore the sky-averaged signal increases during that hour. Further, assume that all the spectra in the first half of the LST bin were excised due to RFI or variability criteria. Then the resulting mean would be biased towards higher sky temperatures and therefore not be comparable to other fully sampled bins at the same LST.
Hence, in order to exclude 1\,h LST bins resulting from averaging spectra with significant gaps, we required that $\tau_{1h}\ge2200$\red{\,s, which corresponds to at least 63 individual 35\,sec integrations, collected in any single day's data.} Bins with smaller integration time were excluded from further analysis.

\item \textbf{Sunrise and sunset}\\
In order to compare only consistent samples of LST bins, we excluded bins that contained either a sunset or a sunrise as indicated in Table~\ref{tab_sun}. It is evident from Table~\ref{tab_sun} that a given 1 h LST bin could be collected during the daytime in one part of the month but during the nighttime at another, due to the ~2\,h drift of \red{LST at} sunrise/sunset during a month. 
We separated these 1\,h LST bins into a daytime set, and a nighttime set, depending on when they occurred. 
Only one of these sets was accepted for further analysis, with the preference given to the nighttime set if it contained at least four bins deemed acceptable according to other criteria.
\end{enumerate}

\begin{table*}
\caption{Sunrise and sunset times at the MRO calculated for start, middle and end of every analyzed month.}
\begin{center}
\begin{tabular}{@{}cccccc@{}}
 Date & \begin{tabular}{@{}c@{}} Sunrise \\ (AWST) \end{tabular} & \begin{tabular}{@{}c@{}} Sunrise \\ (LST) \end{tabular} & \begin{tabular}{@{}c@{}} Sunset \\ (AWST) \end{tabular} & \begin{tabular}{@{}c@{}} Sunset \\ (LST) \end{tabular} & \begin{tabular}{@{}c@{}} LST - AWST(hours) \\ at local midday \end{tabular} \\  
\hline%
2014-10-24 & 05.48 & 07.42 & 18.38 & 20.30 & 1.9 \\
2014-11-01 & 05.37 & 07.83 & 18.48 & 20.91 & 2.4 \\
2014-11-15 & 05.23 & 08.61 & 18.63 & 21.99 & 3.3 \\
2014-11-30 & 05.17 & 09.60 & 18.85 & 23.25 & 4.3 \\
2014-12-15 & 05.22 & 10.56 & 19.02 & 00.33 & 5.3 \\
2014-12-31 & 05.35 & 11.75 & 19.13 & 01.51 & 6.4 \\
2015-01-15 & 05.53 & 12.92 & 19.17 & 02.53 & 7.3 \\
2015-01-31 & 05.75 & 14.18 & 19.10 & 03.51 & 8.4 \\
\hline
\end{tabular}
\end{center}
\label{tab_sun}
\end{table*}

\begin{figure*}
  \begin{center}
    \includegraphics[width=7in]{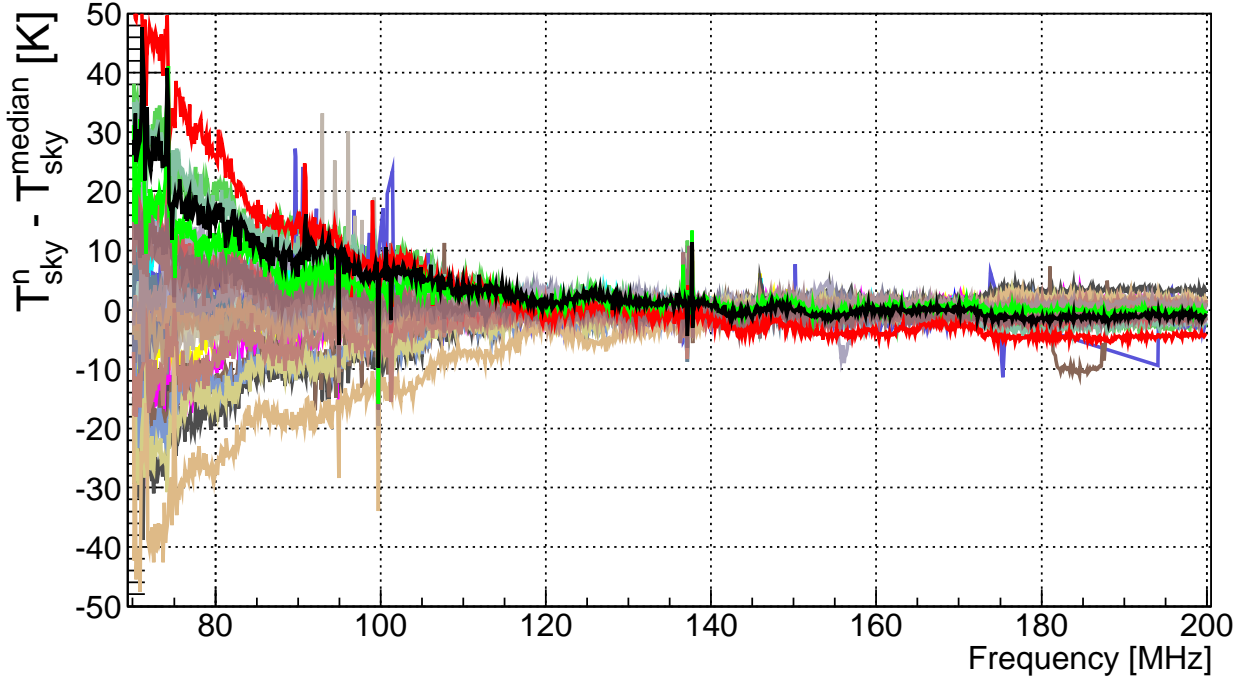}
    \caption{Example of differences calculated between spectra averaged in $0-1$\,h LST ranges and a median spectrum of all 1\,h LST spectra for the data collected between 2014-10-24 and 2014-11-30. \red{The data are presented in the original $117.2$\,kHz resolution and some channels affected by RFI have been removed for clarity.}}
   \label{fig_lst0_differences_nov_oct}
  \end{center}
\end{figure*}

After all the above criteria were applied the number of 1\,h LST bins was reduced from $N_{days}$ to a typically smaller number $N_{acc}$, where $N_{acc}$ varied between a few (for near sunset/sunrise bins) and 32 (for nighttime bins in the Oct/Nov 2014 dataset).
Next, for every 1\,h LST spectrum ($0-1$\,h, $1-2$\,h, $2-3$\,h and so on) a reference sky spectrum was calculated as a median of the $N_{acc}$ spectra accepted for this LST range (Fig.~\ref{fig_1h_dynamic_spectrum} lower figure).
Then for each spectrum in the dynamic spectrum at 1\,h resolution a corresponding reference median spectrum was subtracted. 
\red{The effect of such a subtraction at the original frequency resolution of 117.2\,kHz for $0-1$ LST bin in November/October is presented in Figure~\ref{fig_lst0_differences_nov_oct} where day-to-day variations in received power at the lower frequencies are clearly visible.} The differences in the observed sky-averaged spectrum in relation to possible ionospheric effects will be discussed in the next section.

\begin{figure}
  \begin{center}
    \includegraphics[width=3in]{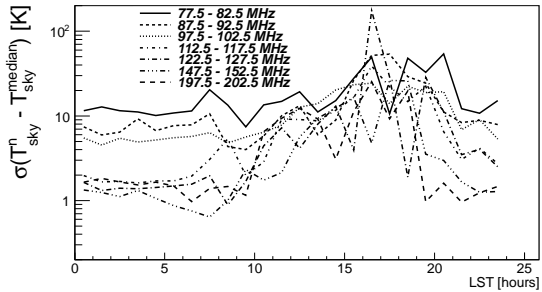}
    \caption{Standard deviation of a Gaussian function fitted to distributions of differences of 1\,h LST spectra and the median spectrum in several frequency bins. The standard deviation is shown as a function of LST. It is stable for 1\,h LST bins collected during the nighttime and is significantly larger for the 1\,h LST bins collected at daytime. The seemingly abrupt transition between $97.5-102.5$ and  $112.5-117.5$ at LST $\le$8\,h results from a steep spectrum of the ionospheric effects at frequencies $\lesssim$110\,MHz (Fig.~\ref{fig_lst0_differences_nov_oct})}.
    \label{fig_rms_vs_lst}
  \end{center}
\end{figure}

\red{
Finally, for the sake of fitting ionospheric parameters, every 40 channels ($\approx$5\,MHz) were averaged and the error of each average channel was calculated as a standard deviation of the individual 40 channels.
The standard deviation of the median-substracted daily signal in several 5\,MHz frequency bins, for all 1\,h LST bins, is shown in Figure~\ref{fig_rms_vs_lst}.
We will further refer to this standard deviation as $\sigma_{iono} \defeq \sigma(T^{n}_{sky} - T^{median}_{sky})$.
Figure~\ref{fig_rms_vs_lst} demonstrates much higher variability in the day-to-day signal at lower frequencies over the range of nighttime LSTs, which we attribute to ionospheric effects.
}


\subsection{Expected ionospheric effects}
\label{subsec_identification_of_iono_effects}

\red{In this subsection we aim to derive an expected frequency dependence, estimate and compare magnitudes of the ionospheric absorption/emission and refraction.}
In the absence of propagation effects due to the ionosphere, the antenna temperature, \red{at a given time $t$}, as a function of frequency resulting from the integration of the sky brightness temperature, $T_{\nu}(\theta,\phi)$,
can be expressed by the following formula (equation 1 in S15):
\begin{equation}
T_{ant}^{ni}(\nu) = \frac{1}{P_{0}} \int_{4\pi} P_{\nu}(\theta,\phi) T_{\nu}(\theta,\phi) d\Omega,
\label{eq_sky_integration_noiono}
\end{equation}
where $P_{\nu}(\theta,\phi)$ is the antenna pattern and the coordinates $(\theta,\phi)$ are angles in the antenna frame, which can be directly translated to horizon coordinates (azimuth, zenith angle) according to orientation of the antenna in the field, and $P_{0} = \int_{4\pi} P_{\nu}(\theta,\phi) d\Omega$ is the normalization factor.
\red{The above formula is time dependent through the implicit time dependence of $T_{\nu}(\theta,\phi)$. Many equations and quantities in this paper are time dependent, but their explicit time dependence is usually not shown for brevity.}

The propagation effects in the ionosphere modify the outside-atmosphere sky brightness temperature $T_{\nu}(\theta,\phi)$ to $T_{\nu}^{p}(\theta,\phi)$ at the antenna, and therefore the antenna temperature observed by a ground-based instrument can be expressed as:
\begin{equation}
T_{ant}(\nu) = \frac{1}{P_{0}} \int_{4\pi} P_{\nu}(\theta,\phi) T_{\nu}^{p}(\theta,\phi) d\Omega.
\label{eq_sky_integration}
\end{equation}
$T_{ant}(\nu)$ is obtained from the calibration procedure (Section~\ref{subsec_initial_data_processing}).
As mentioned in Section~\ref{sec_intro}, the ionosphere contributes most significantly to propagation effects relevant to sky-averaged spectra via absorption, emission, and refraction.

\subsubsection{Contribution from ionospheric absorption and emission}
\label{subsec_contrib_iono_abs}

The ionospheric absorption, \red{at a given time $t$}, can be characterized by \red{the optical depth} $\tau(\nu,\theta,\phi)$ (resulting in signal loss in the ionosphere $L_{iono}(\nu,\theta,\phi) = e^{-\tau(\nu,\theta,\phi)}$) and thermal emission of electrons at temperature $T_e$ (assumed here to be direction independent). 
Therefore, based on radiative transfer \citep{rybicki}, the ionosphere-modified sky brightness temperature, $T_{\nu}^{p}(\theta,\phi)$, can be expressed as:
\begin{equation}
\begin{split}
T_{\nu}^{p}(\theta,\phi) = T_{\nu}(\theta,\phi) e^{-\tau(\nu,\theta,\phi)} + T_e ( 1 - e^{-\tau(\nu,\theta,\phi)} ) \\
\approx T_{\nu}(\theta,\phi) ( 1 - \tau(\nu,\theta,\phi) ) + T_e \tau(\nu,\theta,\phi)
\label{eq_abs_emission}
\end{split}
\end{equation}
where the approximation assumes $\tau \ll 1$.

We further assume that \red{$\tau(\nu,\theta,\phi)$,} is symmetric in azimuth, which due to diurnal gradients in solar irradiation is only an approximation.
\red{We tested this assumption using the slant TEC (STEC) data obtained from the Space Weather Services (SWS) webpage\footnote{\red{\url{http://www.ips.gov.au/World\_Data\_Centre/1/1}}}, which was computed for the MRO location based on GPS satellite data from Australian GPS stations.}
\red{According to these data, the symmetry assumption is valid within $\sim$10-20\% in the North-South direction (slightly better in the East-West direction).}
\red{Moreover, the occurrence of tubular plasma over-density structures aligned along Earth's magnetic field in the upper ionosphere has recently been reported by \citet{cleo_loi}, which occasionally introduce plasma density irregularities of the order of 10\%. Nevertheless, such unusual effects were not taken into account in our estimations.}

Therefore, within \red{our approximation, $\tau(\nu,\theta,\phi)$} can be factorized into a product of a frequency dependent term at zenith $\tau_{z}(\nu)$ and a frequency and azimuth independent geometric \red{term $r_{g}(\theta)$}, which depends only on zenith angle:

\begin{equation}
\begin{split}
\red{ \tau(\nu,\theta,\phi) = \tau_{z}(\nu) r_{g}(\theta) \approx \tau_{100 MHz} \Big(\frac{100}{\nu_{MHz}}\Big)^2 r_{g}(\theta), }
\label{eq_tau_factor}
\end{split}
\end{equation}

\begin{figure}
  \begin{center}
    \includegraphics[width=3in]{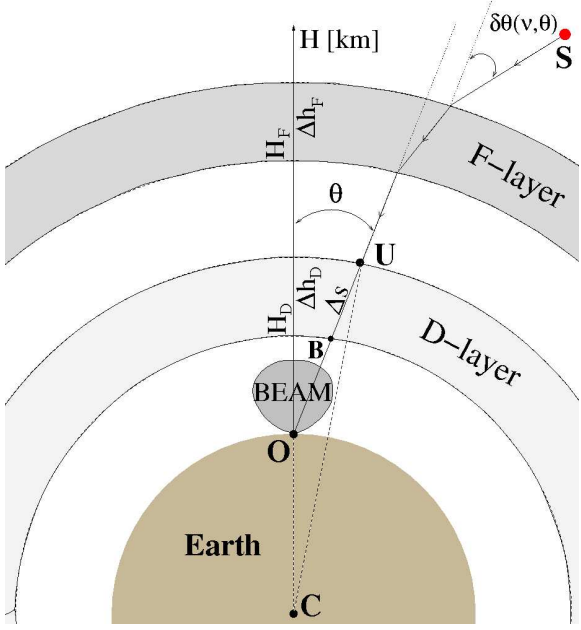}
    \caption{\red{Schematic of the ionospheric refraction in the F-layer and absorption/emission in the D layer. Due to the ionospheric refraction the radio-ray intercepted by the antenna at zenith angle $\theta$ originates from a slightly lower part of the sky at zenith angle $\theta + \delta \theta(\nu,\theta)$.}}
    \label{fig_earth_ionosphere}
  \end{center}  
\end{figure}

\red{where we introduce $\tau_{100 MHz} = 0.23 \times L_{dB}^{100 MHz} \sim 0.0023$ representing \red{optical depth} at zenith at 100\,MHz and $r_{g}(\theta) = \Delta s / \Delta h_D$ is a ratio of ray path length $\Delta s$ in the D-layer to the width $\Delta h_D$ of the D-layer (Fig.~\ref{fig_earth_ionosphere}). The exact equation for $r_{g}(\theta)$, which was derived from geometrical considerations of the triangles $COU$ and $COB$ (Fig.~\ref{fig_earth_ionosphere}) using the law of cosines, can be approximated by the following formula:}

\begin{equation}
\red{r_{g}(\theta) \approx \frac{1 + H_D/R_e}{\sqrt{cos^2(\theta) + 2H_D/R_e}},}
\label{eq_tau_g}
\end{equation}

\red{where $H_D$=75\,km is the height of the D-layer above the surface of the Earth of radius $R_e$=6371\,km, and $\Delta h_D=30$\,km is a mean width of the D-layer.}
\red{Because the length of the radio-ray in the ionosphere depends on its path, the value of $r_{g}(\theta)$ factor is also slightly modified by ionospheric refraction, but we estimated this effect to be negligible as the path length change is $\lesssim$0.1\% even for relatively high refraction effects $\sim$1\,arcmin at 100\,MHz and 45$\degree$ zenith angle.}

The antenna pattern of a conical log spiral antenna should, by design, be frequency independent. 
Agreement between expected antenna properties and the real antenna was confirmed by comparison of calibrated signal and predictions of the sky model integrated with the FEKO-simulated antenna pattern and by the measured reflection coefficient of the conical log spiral antenna in the $70-200$\,MHz frequency range (Fig.~\ref{fig_s11mag_cone_vs_bicon}).
Therefore, in order to simplify our derivation, we further assume a frequency independent antenna pattern, $P_{\nu}(\theta,\phi) = P(\theta,\phi)$.

After substituting equation \ref{eq_tau_factor} into equation \ref{eq_abs_emission} and then equation \ref{eq_abs_emission} into equation~\ref{eq_sky_integration}, \red{we can express the ionosphere-affected antenna temperature $T_{ant}(\nu)$ as a sum of ``no-ionosphere'' antenna temperature $T_{ant}^{ni}(\nu)$ and a ionospheric perturbation term:}

\begin{equation}
\begin{split}
\red{ T_{ant}(\nu) =  T_{ant}^{ni}(\nu) + \tau_{100 MHz} \Big(\frac{100}{\nu_{MHz}}\Big)^2 \frac{1}{P_{0}} \times} \\
\red{  \bigg( \int_{4\pi} P_{\nu}(\theta,\phi) T_e  r_{g}(\theta)d\Omega - 
      \int_{4\pi} P_{\nu}(\theta,\phi) T_{\nu}(\theta,\phi) r_{g}(\theta)d\Omega \bigg) }.
\label{eq_long}
\end{split}
\end{equation}

\red{The second integral can be expressed based on the mean value theorem for integrals as:}

\begin{equation}
\begin{split}
\red{\frac{1}{P_{0}} \int_{4\pi} P_{\nu}(\theta,\phi) T_{\nu}(\theta,\phi) r_{g}(\theta)d\Omega = }\\
\red{r_{g}\prime \, \frac{1}{P_{0}} \int_{4\pi} P_{\nu}(\theta,\phi) T_{\nu}(\theta,\phi) d\Omega = r_{g}\prime \; T_{ant}^{ni}(\nu),}
\label{eq_mean_value_theorem}
\end{split}
\end{equation}

\red{where $T_{ant}^{ni}(\nu)$ was substituted according to equation~\ref{eq_sky_integration_noiono} and $r_{g}\prime$ is the $P_{\nu}(\theta,\phi) T_{\nu}(\theta,\phi)$-weighted average of $r_{g}(\theta)$. After defining $r_{g} = \frac{1}{P_{0}} \int_{4\pi} P_{\nu}(\theta,\phi) r_{g}(\theta)d\Omega$, equation~\ref{eq_long} can be expressed as:}

\begin{equation}
\begin{split}
\red{T_{ant}(\nu)  \approx T_{ant}^{ni}(\nu) + \tau_{100 MHz} T_e \Big(\frac{100}{\nu_{MHz}}\Big)^2 r_g} \\
\red{                     - \tau_{100 MHz} \Big(\frac{100}{\nu_{MHz}}\Big)^2 T_{ant}^{ni}(\nu) r_g\prime}
\label{eq_long_approx}
\end{split}
\end{equation}

\red{Based on the FEKO simulated antenna pattern, the value of the frequency independent geometric factor $r_g$ was found to be $r^{cone}_{g}\approx$\,1.4 and, as expected, almost constant at frequencies above 75\,MHz.}
\red{For comparison we have also calculated $r^{gauss}_g \approx 1.7$ for a Gaussian symmetrical beam ($\sim exp(-(\theta/52\degree)^2)$), which we found to be a good and smooth approximation of the FEKO simulated pattern of the conical log spiral antenna (Fig~\ref{fig_cone_pattern_feko}).}
\red{Similarly, the value of frequency independent geometric factor $r_g\prime$ for the cone was found to be $r_g\prime \approx$\,1.48.}

The contribution from ionospheric absorption and emission to the difference between the spectrum $T_{\nu}^n(\nu)$ collected on the $n^{th}$ day and a reference spectrum $T_{\nu}^r(\nu)$, both collected in the same LST range but at different ionospheric conditions characterized by two pairs of parameters $(\tau_{100 MHz}^n,T_e^n)$ and $(\tau_{100 MHz}^r,T_e^r)$, can be expressed by the following formula:

\begin{equation}
\begin{split}   
\Delta_{nr}^{ae}(\nu) = T_{ant}^n(\nu) - T_{ant}^r(\nu) = \\
\red{(\tau_{100 MHz}^r - \tau_{100 MHz}^n) \Big(\frac{100}{\nu_{MHz}}\Big)^2 T_{ant}^{ni}(\nu) r_{g}\prime +} \\
\red{(\tau_{100 MHz}^n T_e^n - \tau_{100 MHz}^r T_e^r) \Big(\frac{100}{\nu_{MHz}}\Big)^2 r_{g}.}
\label{eq_diff_spectra}
\end{split}
\end{equation}

\red{If we approximate the sky temperature with a power-law \red{$T_{ant}^{ni}(\nu) \approx T_{100} (100/\nu_{MHz})^\alpha$, where $\alpha \approx 2.6$ \citep{2008AJ....136..641R}} and $T_{100}$ is the sky temperature at 100\,MHz, }
we obtain the expected frequency dependence of the difference between two spectra collected at the same LST expressed as a \red{sum of two power-laws $\sim1/\nu^2$ and $\sim1/\nu^{2+\alpha}$:}

\begin{equation}
\begin{split}
\red{\Delta_{nr}^{ae}(\nu) \approx r_{g} \Delta_{T\tau} \Big(\frac{100}{\nu_{MHz}}\Big)^2 + r_{g}\prime \Delta_\tau T_{100} \Big(\frac{100}{\nu_{MHz}}\Big)^{2+\alpha},}
\label{eq_diff_spectra_simple}
\end{split}
\end{equation}

where we define $\Delta_\tau = (\tau_{100 MHz}^r - \tau_{100 MHz}^n)$ and $\Delta_{T\tau} = (\tau_{100 MHz}^n T_e^n - \tau_{100 MHz}^r T_e^r)$.

\red{Because the values of $r_g$ and $r_g\prime$ are very similar we further approximated $r_g\prime \approx r_g \approx 1.4$, which simplifies the following analysis without impacting the frequency dependence. Nevertheless, this approximation has a small ($\lesssim$5\%) impact on the measured electron temperatures.}

\subsubsection{Contribution from ionospheric refraction}
\label{subsec_contrib_iono_refr}

%
%
%

\red{Ionospheric refraction is relevant to global EoR experiments because it distorts the sky observed by a ground-based instrument such that the ionosphere acts as a huge slightly de-magnifying lens. Therefore, from the antenna perspective, a certain pointing $(\theta,\phi)$ within the antenna beam gets deflected by the ionosphere such that the antenna intercepts the radiation originating from a slightly lower sky position $(\theta+\delta \theta(\nu,\theta),\phi)$ (Fig.~\ref{fig_earth_ionosphere}).}

\red{As described in Section~\ref{subsec_ionosphere}, the typical mean offset expected from the theoretical formula by \citet{bailey_1948} is $\delta \theta_{0} \sim0.1$\,arcmin at 100\,MHz and zenith angle 45$\degree$ and varies proportionally to variations in electron density.}
\red{Similarly, based on the data from the MWA telescope (co-located with our instrument) \citet{balwinder} report nighttime offsets of the source positions of the order of tenths of arcmin at 100\,MHz and scaling as $1/\nu^2$.}

\red{Therefore, based on the TEC values computed from the GPS data for the MRO (the same data from the SWS webpage as in Sec.~\ref{subsec_contrib_iono_abs}), we estimate that the intra-day variations of the reference offset $\delta \theta_{0}$ within a month should typically be $\sim$20-30\% and can be higher during the periods of higher solar activity.}
\red{Bailey's formula for the deflection angle can be expressed as a product of frequency and zenith angle dependent terms as:}

\begin{equation}
\red{\delta\theta (\nu,\theta) = \delta\theta(\nu) \delta \theta(\theta) \approx \delta\theta_{0}(100/\nu_{MHz})^2 \frac{\delta(\theta)}{\delta(45\degree)},}
\label{eq_delta_theta}
\end{equation}

\red{where $\delta\theta_{0} \sim 0.1$\,arcmin. The exact form of frequency independent geometric factor $f(\theta)$ can be found in \citep{bailey_1948,datta_et_al}.}
\red{After including refraction, equation~\ref{eq_sky_integration_noiono} can be re-written as:}

\begin{equation}
\red{T_{ant}^{refr}(\nu) = \frac{1}{P_{0}} \int_{4\pi} P_{\nu}(\theta,\phi) T_{\nu}(\theta + \delta \theta(\nu,\theta),\phi) d\Omega,}
\label{eq_sky_integration_ionorefr}
\end{equation}

\red{where $\delta \theta(\nu,\theta)$ can be calculated from equation~\ref{eq_delta_theta}.}
\red{In order to estimate the magnitude of refraction, we implemented the above equation in the modeling software and calculated sky-averaged spectra for the ``non-ionospheric'' case ($T_{ant}^{ni}(\nu)$) and for several refraction cases ($T_{ant}^{refr}(\nu)$) represented by different values of $\delta\theta_{0}$ ranging from 0.1\,arcmin (typical nighttime) to 5\,arcmin (extreme daytime case). Next, we calculated differences between these sky-averaged spectra. }
\red{The differences between spectra affected by typical nighttime refraction ($\delta\theta_{0}=0.1$\,arcmin) and the hypothetical ``non-ionospheric'' scenario, compared with similar differences for absorption/emission effects are shown in Figure~\ref{fig_ionoeffect_minus_noeffect}.}
\red{The contribution of nighttime refraction to the sky-averaged spectrum above 80\,MHz is $\lesssim$1\,K, which is much less than the expected contribution of absorption and emission.}

\red{We also calculated differences between sky-averaged spectra affected by different refraction conditions ($\delta\theta_{0} = 0.1 - 0.3$\,arcmin) and found out that even relatively large ($\sim$0.1\,arcmin) variations of $\delta\theta_{0}$ around a mean value lead to a very small ($\lesssim$1\,K) changes of the sky-averaged spectra.}
\red{However, the same tests for daytime conditions (with $\delta\theta_{0}$ in the range $1 - 5$\,arcmin) show that during daytime contribution of the ionospheric refraction may be of the order of tens of Kelvins.}

\red{In order to derive the frequency dependence of the ionospheric refraction, equation~\ref{eq_sky_integration_ionorefr} can be approximated by the first two terms of its Taylor expansion:}
\begin{equation}
\begin{split}
\red{T_{ant}^{refr}(\nu) = \frac{1}{P_{0}} \int_{4\pi} P_{\nu}(\theta,\phi) T_{\nu}(\theta,\phi) d\Omega  + }\\ 
\red{\delta \theta (\nu) \frac{1}{P_{0}} \int_{4\pi} P_{\nu}(\theta,\phi) \frac{dT_{\nu}(\theta,\phi)}{d\theta} \delta \theta (\theta) d\Omega,}
\label{eq_sky_integration_ionorefr_taylor}
\end{split}
\end{equation}

\red{where the first term is the antenna temperature $T_{ant}^{ni}(\nu)$ defined in equation~\ref{eq_sky_integration_noiono} as measured without (or outside) the ionosphere and we expressed $\delta \theta(\nu,\theta)$ as a product of frequency and angle dependent terms (Eq.~\ref{eq_delta_theta}).}
\red{It is worth noting that the second term vanishes when $dT_{\nu}(\theta,\phi)/ d\theta = 0$, meaning that refraction effects vanish in a hypothetical ``constant sky'' scenario (i.e. when $T_{\nu}(\theta,\phi)$ is constant).}

\red{Assuming, as before, a power-law dependence of the sky temperature $T_{\nu}(\theta,\phi) = T_{100}(\theta,\phi) (100 / \nu_{MHz})^\alpha$ equation~\ref{eq_diff_refr} can be approximated by:}
\begin{equation}
\begin{split}
\red{T_{ant}^{refr}(\nu) \approx T_{ant}^{ni}(\nu) + \delta \theta (\nu) \Big(\frac{100}{\nu_{MHz}}\Big)^\alpha I_{dT},}
\label{eq_sky_integration_ionorefr_taylor2}
\end{split}
\end{equation}
\red{where we define a frequency independent integral $I_{dT}$ as}
\begin{equation}
\begin{split}
\red{I_{dT} = \frac{1}{P_{0}} \int_{4\pi} P_{\nu}(\theta,\phi) \frac{dT_{100}(\theta,\phi)}{d\theta} \delta \theta (\theta) d\Omega.}
\label{eq_dp}
\end{split}
\end{equation}

Hence, the refraction induced difference $\Delta_{nr}^{refr}(\nu)$ between the spectrum $T_{\nu}^{refr,n}(\nu)$ collected on the $n^{th}$ day and a reference spectrum $T_{\nu}^{refr,r}(\nu)$, both collected at the same LST but at different ionospheric conditions resulting in different refraction angles $\delta\theta^n$ and $\delta\theta^r$, can be expressed by the following formula:
\begin{equation}
\begin{split}
\red{\Delta_{nr}^{refr}(\nu) = (\delta\theta^n(\nu) - \delta\theta^r(\nu)) \Big(\frac{100}{\nu_{MHz}}\Big)^\alpha I_{dT}}\\
\red{\approx (\delta\theta_{0}^{n} - \delta\theta_{0}^{r}) \Big(\frac{100}{\nu_{MHz}}\Big)^{2+\alpha} I_{dT},}
\label{eq_diff_refr}
\end{split}
\end{equation}

\red{where we used equation~\ref{eq_delta_theta} to expand frequency dependence of $\delta\theta^n(\nu)$ and $\delta\theta^r(\nu)$.}


The exponent $(2+\alpha)$ in equation~\ref{eq_diff_refr} is exactly the same as the exponent in one of the components in equation~\ref{eq_diff_spectra_simple}.
Hence, we conclude that the difference $\Delta_{nr}(\nu) = \Delta_{nr}^{ae}(\nu) + \Delta_{nr}^{refr}(\nu)$ of two spectra $n$ and $r$ collected at the same LST, due to ionospheric absorption, emission and refraction, can be described by a sum of power-laws:
\begin{equation}
\begin{split}
\Delta_{nr}(\nu) = A (100/\nu_{MHz})^2 + B (100/\nu_{MHz})^2 \, T_{ant}(\nu) \\
\approx A (100/\nu_{MHz})^2 + B' (100/\nu_{MHz})^{2+\alpha}.
\label{eq_diff_total}
\end{split}
\end{equation}

\subsection{Comparison of expected ionospheric contributions}
\label{subsec_comparison_of_iono_contribs}

In order to compare contributions of the ionospheric effects, they were incorporated into the signal modeling software described in S15. Ionospheric absorption and emission were included according to equations~\ref{eq_sky_integration},~\ref{eq_abs_emission}~and~\ref{eq_tau_factor}, whilst refraction was included according to \red{equation~\ref{eq_sky_integration_ionorefr}}.

\begin{figure}
  \begin{center}
    \includegraphics[width=3in]{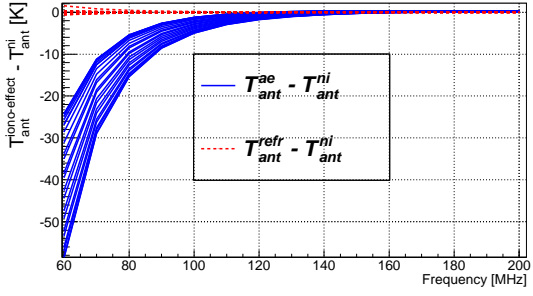}
    \caption{\red{Example simulations of differences of sky-averaged spectra affected and unaffected by ionospheric effects generated for 24\,hours of local time (different curves represent different LSTs).
The dashed red lines represent the contribution from ionospheric refraction (equations~\ref{eq_sky_integration_ionorefr} and \ref{eq_delta_theta}) and solid blue lines represent the overall change of received power due to ionospheric absorption and emission (equations~\ref{eq_sky_integration},~\ref{eq_abs_emission},~\ref{eq_tau_factor} with $L_{dB}^{100 MHz}$ = 0.01\,dB).}}
    \label{fig_ionoeffect_minus_noeffect}
  \end{center}  
\end{figure}

Because the antenna gain values were calculated by the FEKO software in discrete directions (every 1\,$\degree$), for the sake of this comparison we used the Gaussian beam described above.
The sky model was integrated with the Gaussian beam in the absence of any ionospheric effects ($T_{ant}^{ni}(\nu)$), in the presence of absorption and emission ($T_{ant}^{ae}(\nu)$), and in the presence of refraction only ($T_{ant}^{refr}(\nu)$).
The expected contribution from absorption/emission was calculated as $\Delta_{model}^{ae}(\nu) = T_{ant}^{ae}(\nu) - T_{ant}^{ni}(\nu)$ and the expected contribution from refraction as $\Delta_{model}^{refr}(\nu) = T_{ant}^{refr}(\nu) - T_{ant}^{ni}(\nu)$ (Fig.~\ref{fig_ionoeffect_minus_noeffect}).

\red{Similarly, we also calculated differences between typical and perturbed parameters (by amounts expected from the TEC variations at the MRO) and the results also support the assertion that absorption and emission will dominate at these frequencies, which is also in agreement with simulations by \citet{harish_et_al}.}
\red{During the nighttime, the contribution from refraction at 80\,MHz is almost negligible (below 1\,K) in comparison to absorption and emission when mean offset and its variations at 100\,MHz and zenith angle 45$\degree$ are $\sim$0.1\,arcmin and slightly larger ($\lesssim$5\,K) for $\sim$0.5-1\,arcmin.}
\red{Thus, we expect that at least during the nighttime, fitting of differences of the spectra with equation~\ref{eq_diff_total} will result in values of parameters A and B being dominated by absorption and emission with a negligible contribution from refraction.}

%
%


\subsection{Fitting ionospheric parameters}
\label{subsec_fitting_proc}

\begin{figure}
  \begin{center}
	 \includegraphics[width=3.4in,height=8.23in]{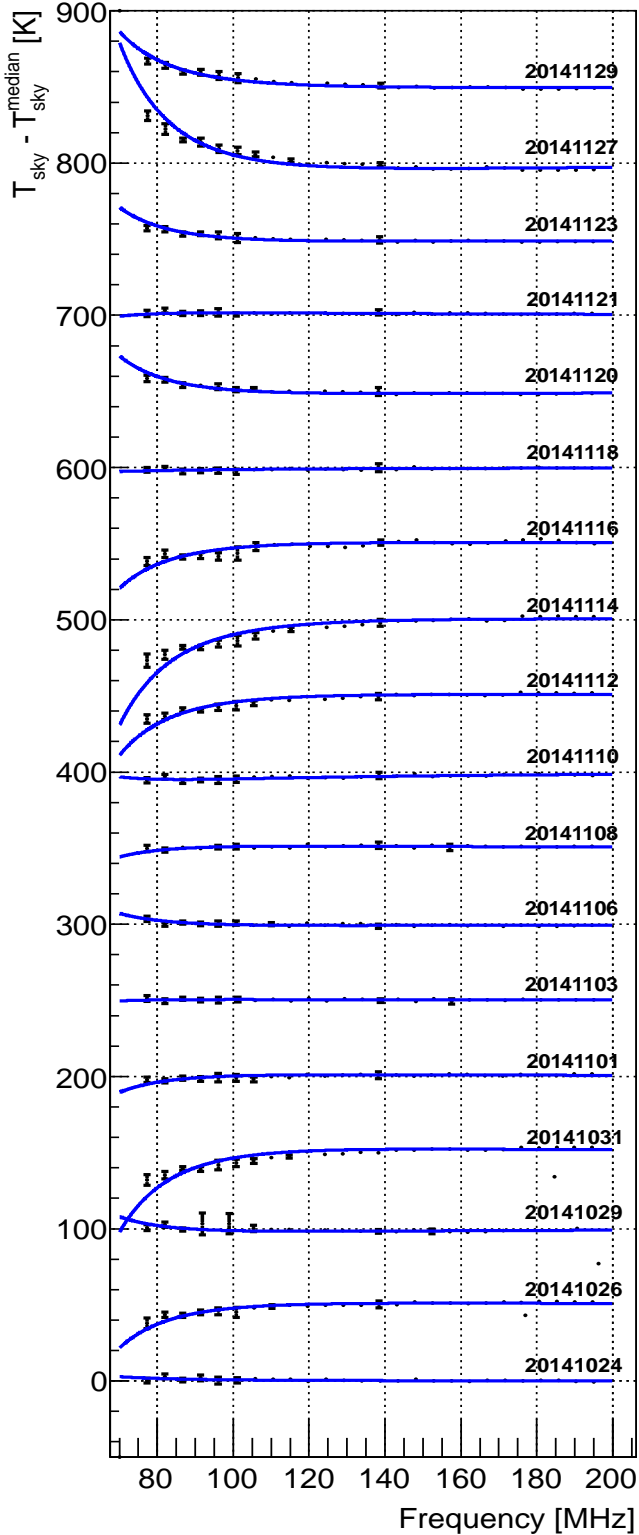}
    \caption{\red{Example results of fitting equation~\ref{eq_diff_total} to differences of $0-1$h LST spectra (as in Figure~\ref{fig_lst0_differences_nov_oct} after averaging of 40 frequency channels) and a reference median spectrum collected in Oct and Nov 2014. The spectra are offset by 50\,K for clarity, but they all lie around zero. Large errors are usually due to residual RFI (for example due to the FM-band in the case of two points in $90-100$\,MHz band on the curve at $\approx$100\,K). The values of fitted parameters A and B, their errors, and reduced $\chi^2$ are listed in Table~\ref{tab_ab_fit_results}.}}
    \label{fig_ab_fit_results}
  \end{center}
\end{figure}

\red{Assuming that the differences between each 1\,h spectrum and the median spectrum (Fig.~\ref{fig_lst0_differences_nov_oct}) are due to variations of the sky signal caused by the ionospheric effects described in Section~\ref{subsec_identification_of_iono_effects}, we  fitted the difference spectra with equation~\ref{eq_diff_total}.}
Example fitting results for the LST $0-1$\,h bin collected in Oct and Nov 2014 are shown in Figure~\ref{fig_ab_fit_results} and Table~\ref{tab_ab_fit_results}.

\subsubsection{Derivation of electron temperature}
\label{subsubsec_te_fits}

\begin{figure*}
\begin{center}
\subfloat[LST = 0 - 1 hours]{\includegraphics[width = 3.6in]{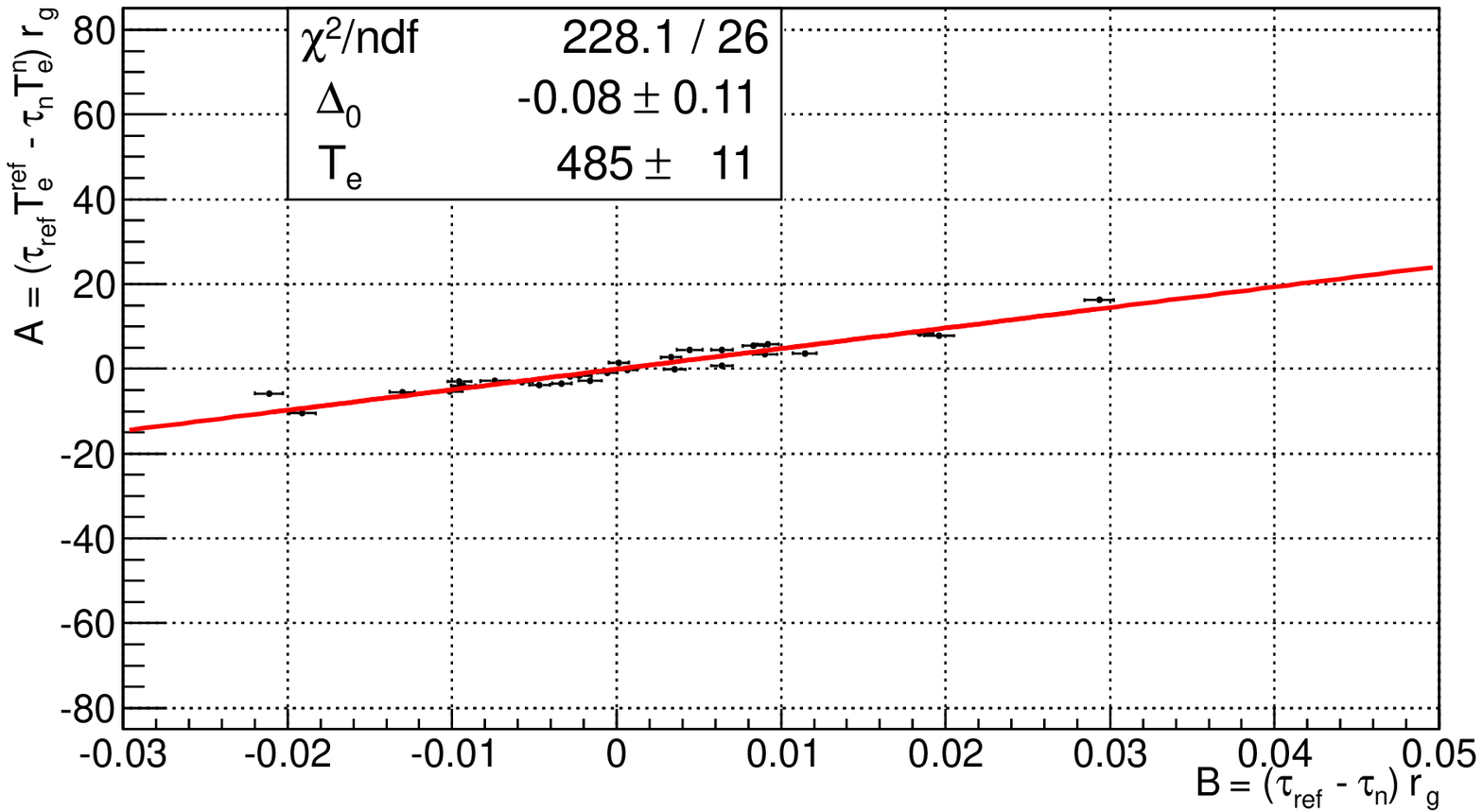}} 
\subfloat[LST = 4 - 5 hours]{\includegraphics[width = 3.6in]{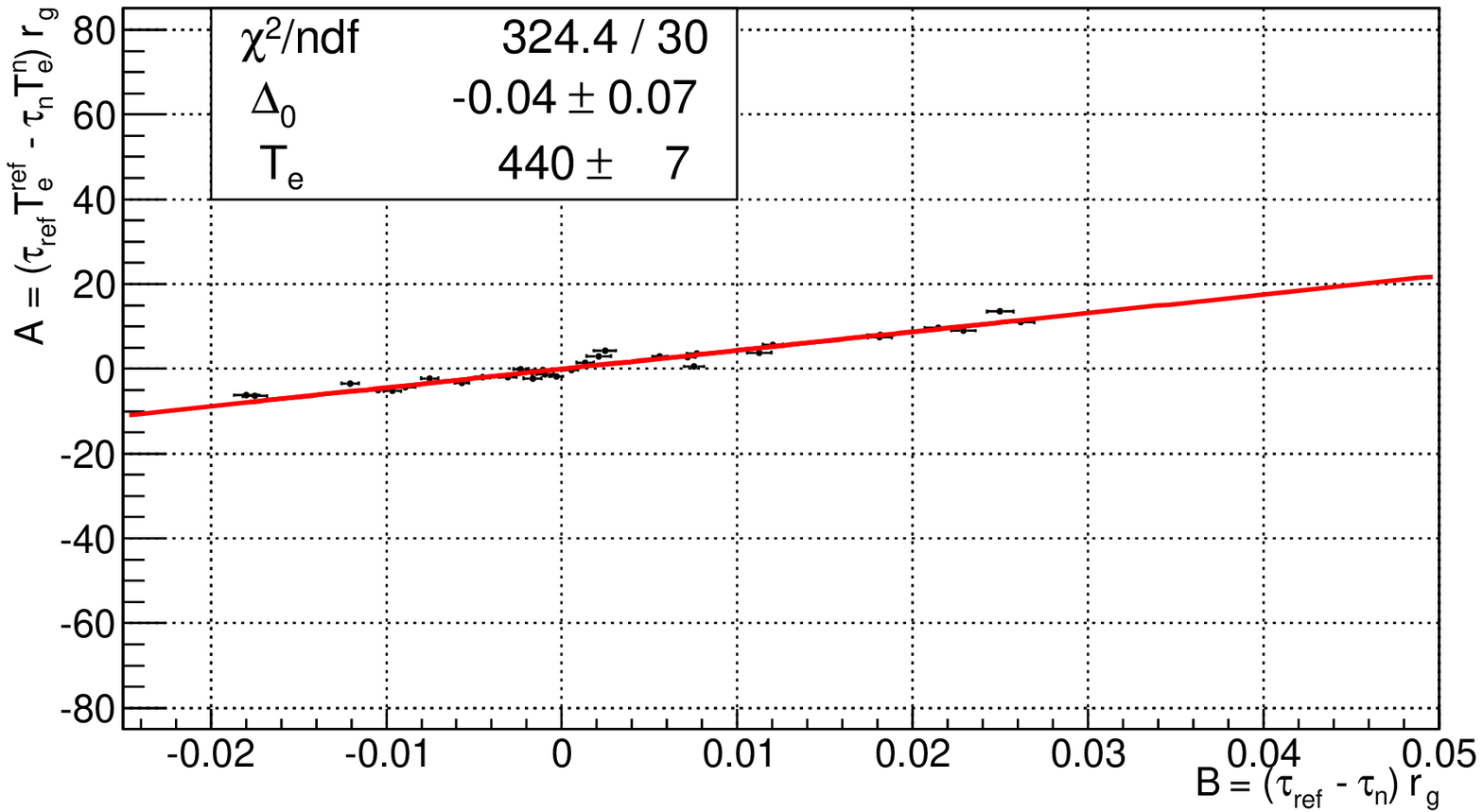}}\\
\subfloat[LST = 8 - 9 hours]{\includegraphics[width = 3.6in]{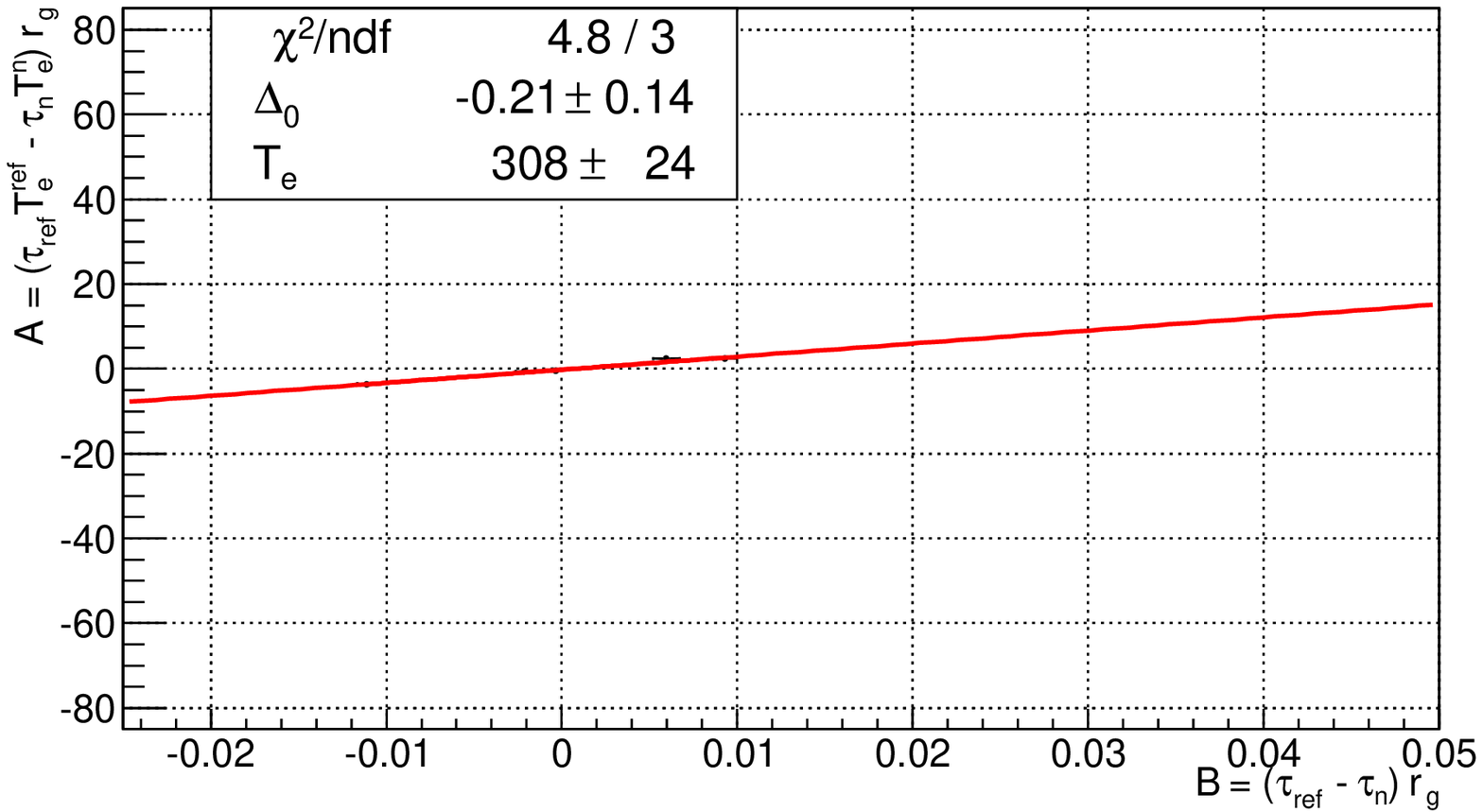}}
\subfloat[LST = 12 - 13 hours]{\includegraphics[width = 3.6in]{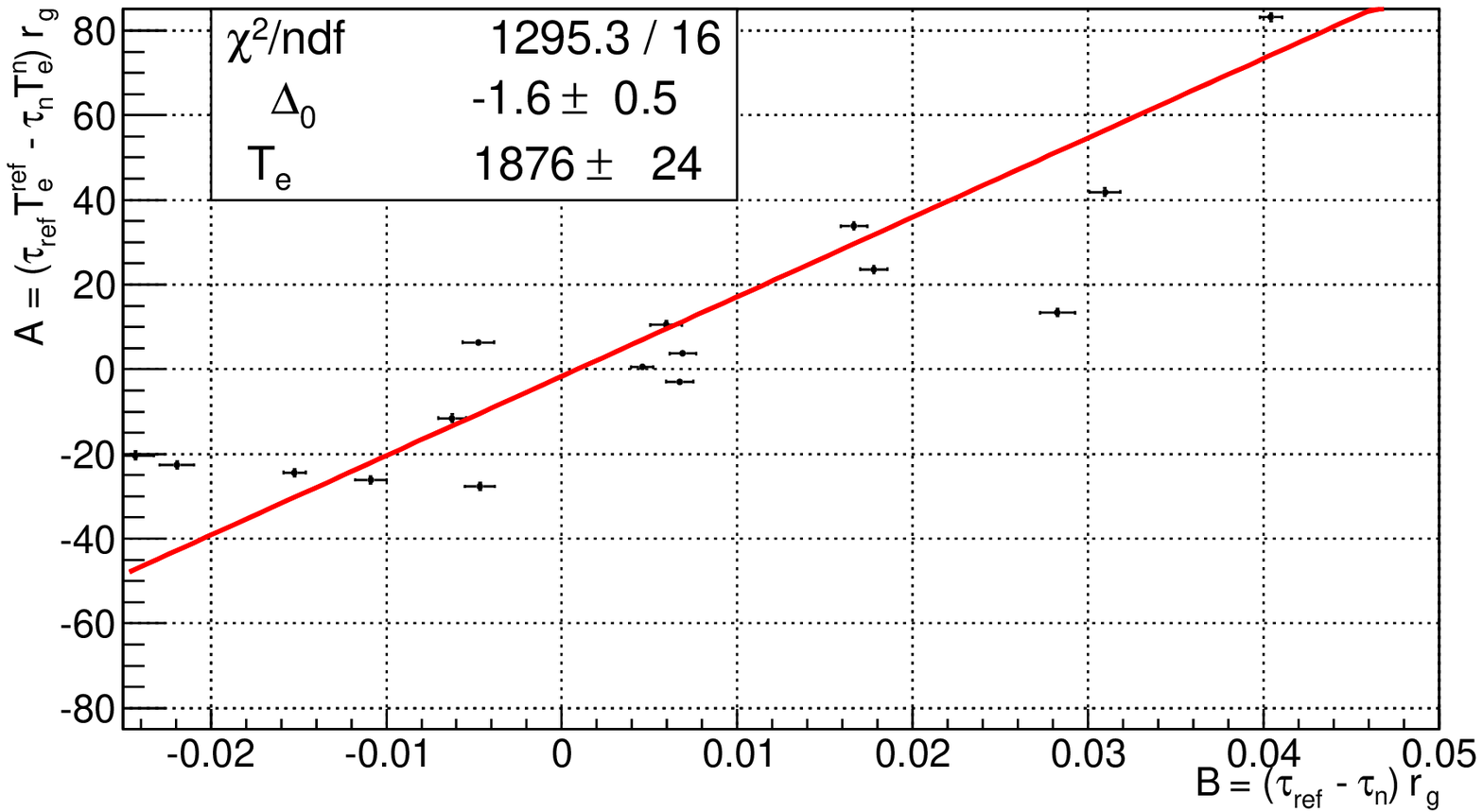}}\\
\subfloat[LST = 16 - 17 hours]{\includegraphics[width = 3.6in]{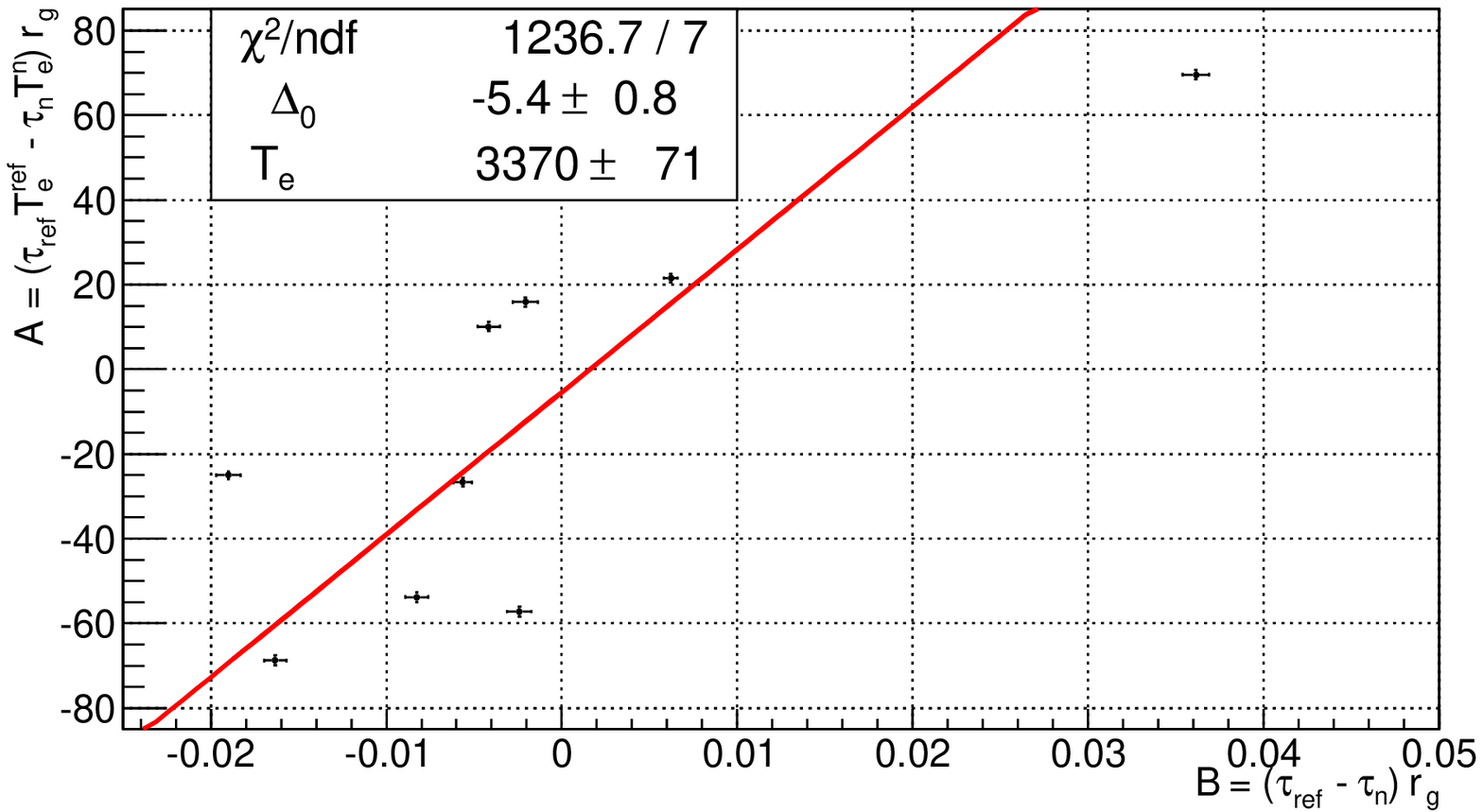}}
\subfloat[LST = 19 - 20 hours]{\includegraphics[width = 3.6in]{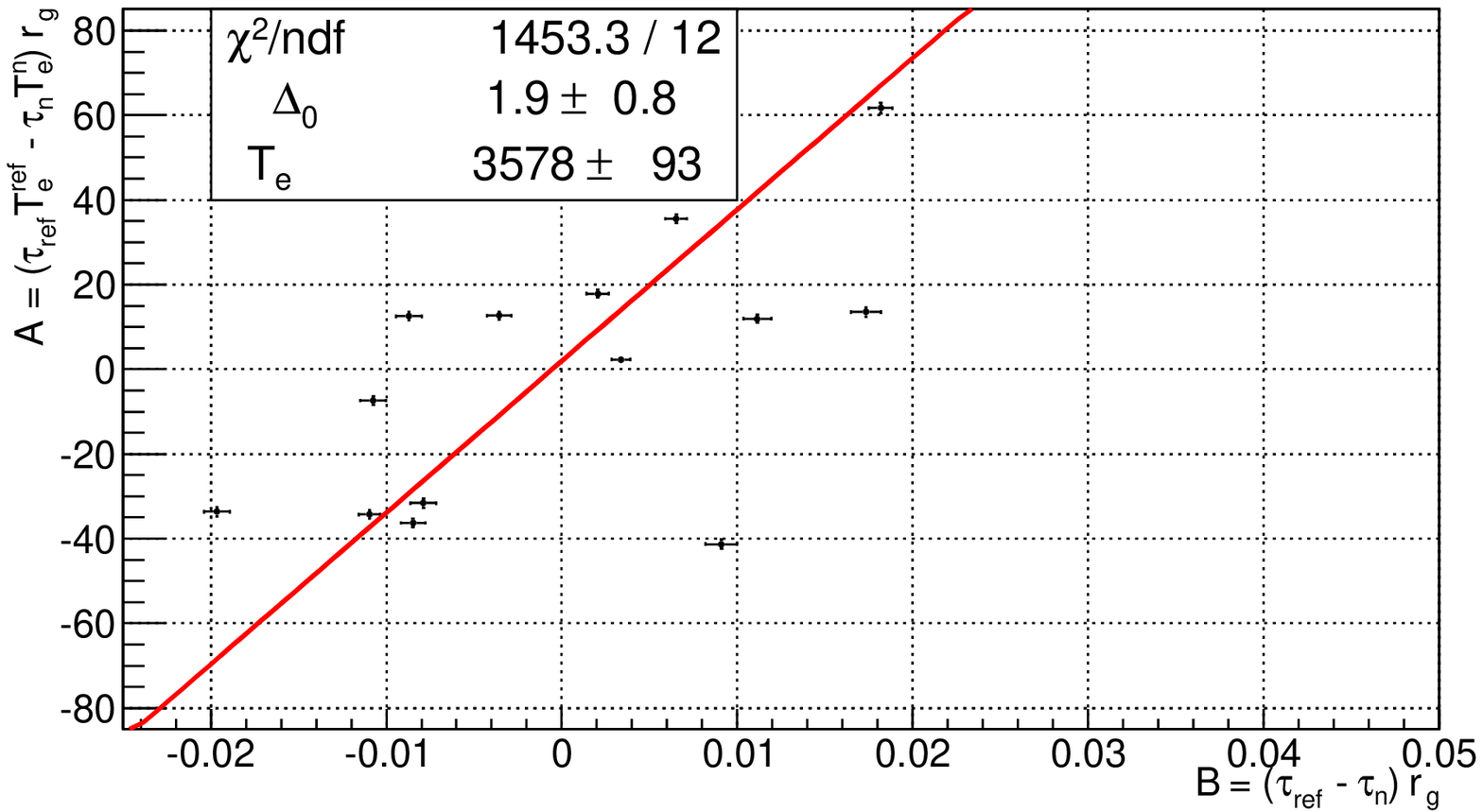}}\\ 
\subfloat[LST = 20 - 21 hours]{\includegraphics[width = 3.6in]{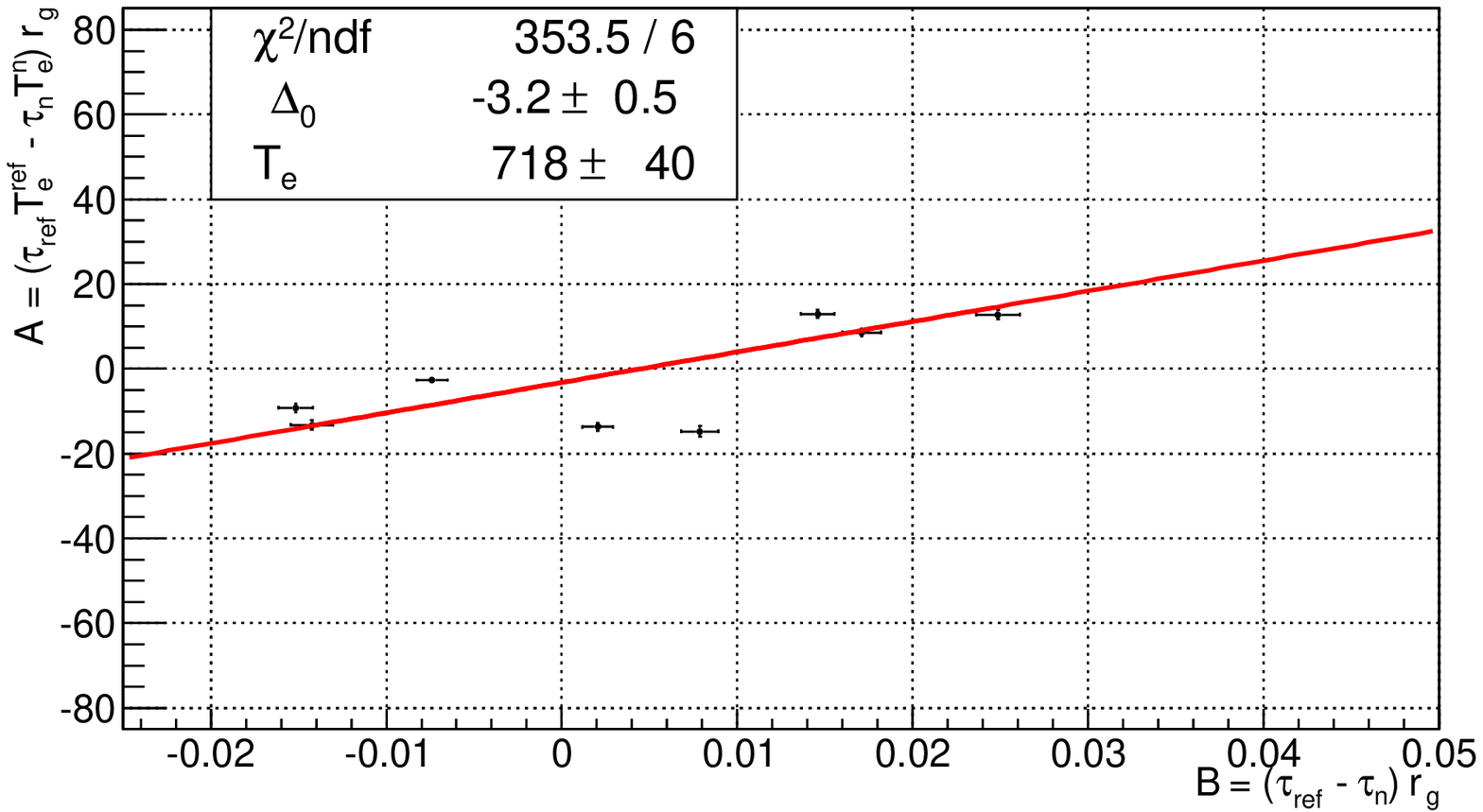}}
\subfloat[LST = 23 - 24 hours]{\includegraphics[width = 3.6in]{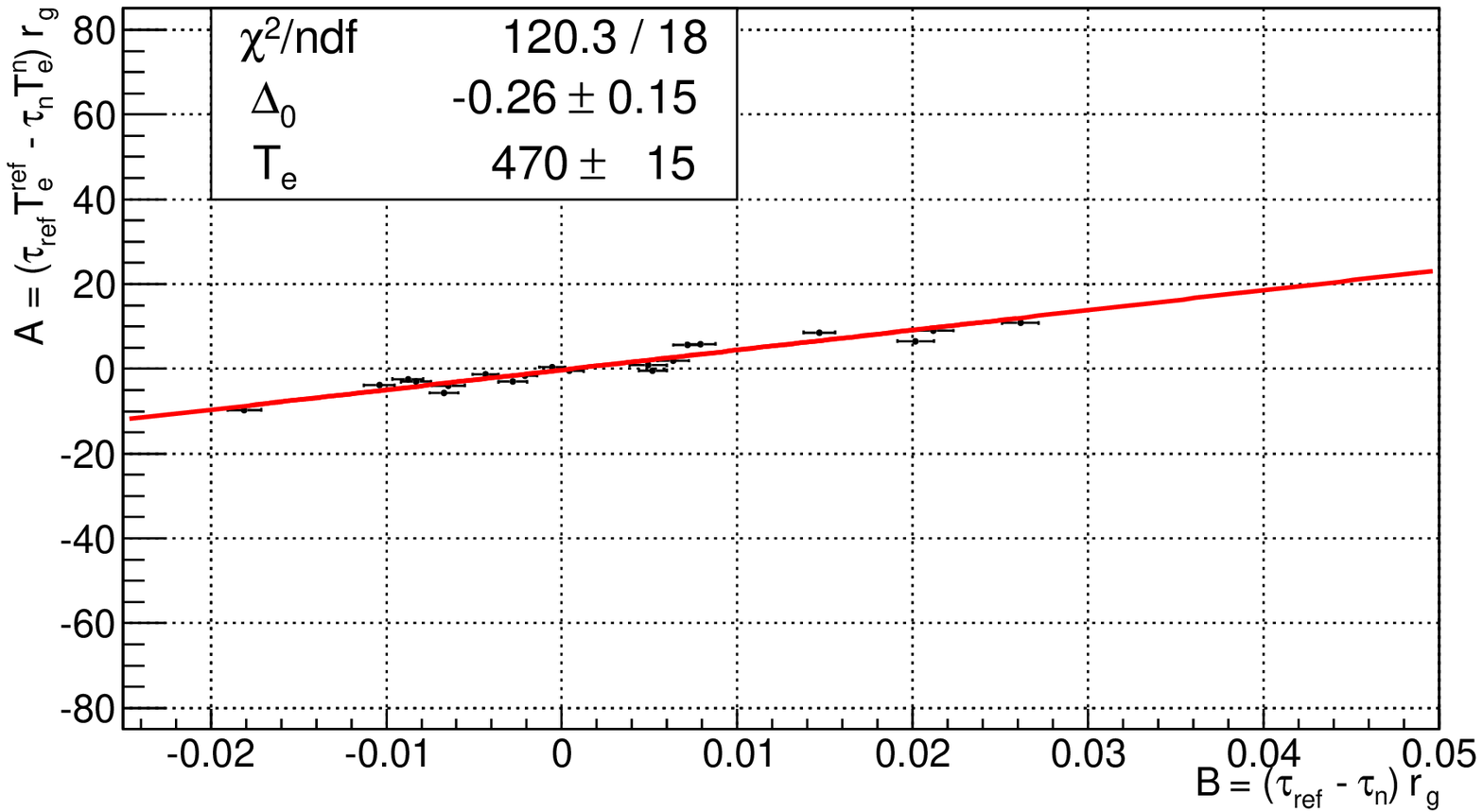}} 
\caption{An example of correlation between fitted parameters A and B for several 1\,h LST bins for the data collected in Oct/Nov 2014. Parameters A and B can be interpreted as in equation~\ref{eq_fitted_ab_interpretation} and fitted values of $T_e$ are shown in each figure. An approximate range in local time can be obtained by subtracting $\approx$3.5\,hours. The small number of data points in figures c) and g) is a result of exclusion of 1\,h LST bins including sunrise and sunset or containing nighttime and daytime spectra.}
\label{fig_a_vs_b}
\end{center}
\end{figure*}

We plotted the fitted A values as a function of B values (eq.~\ref{eq_diff_total}); an example of their dependence for the $0-1$\,h LST bin is shown in Figure~\ref{fig_a_vs_b}, which suggests a linear correlation between these parameters (at least at some LSTs).
The fitting procedure was repeated for all LST ranges and the results also exhibit a very similar linear correlation, particularly for the nighttime data.
Based on equation~\ref{eq_diff_spectra_simple}, the fitted parameters A and B can be interpreted as a combination of frequency independent parameters
\begin{equation}
\begin{split}
\red{B = (\tau_{100 MHz}^r - \tau_{100 MHz}^n) r_{g}\prime, }\\
\\
\red{A = (\tau_{100 MHz}^n T_e^n - \tau_{100 MHz}^r T_e^r) r_{g} }\\
\red{\approx T_e (\tau_{100 MHz}^n  - \tau_{100 MHz}^r ) r_{g} \approx T_e B,}
\label{eq_fitted_ab_interpretation}
\end{split}
\end{equation}
where we assumed that for a given 1\,h LST range the electron temperature is approximately constant $T_e^n \approx T_e^r \approx T_e$ \red{and $r_{g}\prime \approx r_{g}$ (see Sec.~\ref{subsec_contrib_iono_abs}).}
Hence, in this approximation $A \approx T_e B$ and therefore the slope fitted to the data in Figure~\ref{fig_a_vs_b} can be interpreted as electron temperature for \red{a given LST range}.

\begin{table}
\caption{\red{Parameters A and B, their errors, and reduced $\chi^2$ resulting from fitting equation~\ref{eq_diff_total} to the data in Figure~\ref{fig_ab_fit_results}.}}
\begin{center}
\begin{tabular}{@{}cccccc@{}}
 Date & A & $\Delta$A & B & $\Delta$B & $\chi^2$/NDF \\
\hline
20141129 &  0.0114  & 0.0007 &  3.59 & 0.41 & 60.7 / 24 \\
20141127 &  0.0293  & 0.0009 & 16.21 & 0.48 & 207.7 / 24 \\
20141123 &  0.0083  & 0.0007 &  5.38 & 0.39 & 32.2 / 24 \\
20141121 & -0.0017  & 0.0007 & -2.89 & 0.41 & 17.1 / 24 \\
20141120 &  0.0091  & 0.0007 &  5.73 & 0.40 & 32.0 / 24 \\
20141118 &  0.0001  & 0.0006 &  1.40 & 0.37 & 14.0 / 24 \\
20141116 & -0.0100  & 0.0008 & -4.51 & 0.46 & 151.6 / 24 \\
20141114 & -0.0212  & 0.0009 & -5.80 & 0.47 & 186.9 / 24 \\
20141112 & -0.0131  & 0.0007 & -5.56 & 0.43 & 79.5 / 24 \\
20141110 &  0.0026  & 0.0006 &  6.51 & 0.39 & 27.0 / 24 \\
20141108 & -0.0034  & 0.0006 & -3.47 & 0.42 & 20.7 / 24 \\
20141106 &  0.0033  & 0.0006 &  2.65 & 0.40 & 34.3 / 24 \\
20141103 & -0.0006  & 0.0006 & -0.99 & 0.40 & 15.5 / 24 \\
20141101 & -0.0047  & 0.0006 & -3.81 & 0.39 & 39.1 / 24 \\
20141031 & -0.0193  & 0.0008 & -10.63 & 0.45 & 158.6 / 24 \\
20141029 &  0.0045  & 0.0008 &  4.54 & 0.54 & 21.1 / 22 \\
20141026 & -0.0102  & 0.0008 & -5.35 & 0.41 & 67.1 / 24 \\
20141024 &  0.0006  & 0.0006 & -0.29 & 0.36 & 18.9 / 24 \\
\hline
\end{tabular}
\end{center}
\label{tab_ab_fit_results}
\end{table}

As further results will show, the constant $T_e$ assumption is valid for the exclusively nighttime 1\,h LST bins, but it is invalid for other LST bins.
\red{Because LST drifts with respect to local time by $\approx$2\,h/month, a 1\,h LST bin effectively covers $\approx$3\,h in local time (the width of the 1\,h LST bin plus 2\,h of the drift between LST and local time) over the period of a month (3.5\,h in the case of Oct/Nov dataset).}
The assumption is invalid particularly for the LST bins near sunrise/sunset when electron temperature changes rapidly (but smoothly) or for the daytime bins when electron temperatures can vary between days.
\red{The results for these LST bins therefore represent an average electron temperature in a 3\,h (3.5\,h in the case of Oct/Nov data)} local time bin (\red{as indicated later by the bars on the time axis}).

\begin{figure}
  \begin{center}
	 \includegraphics[width=3in]{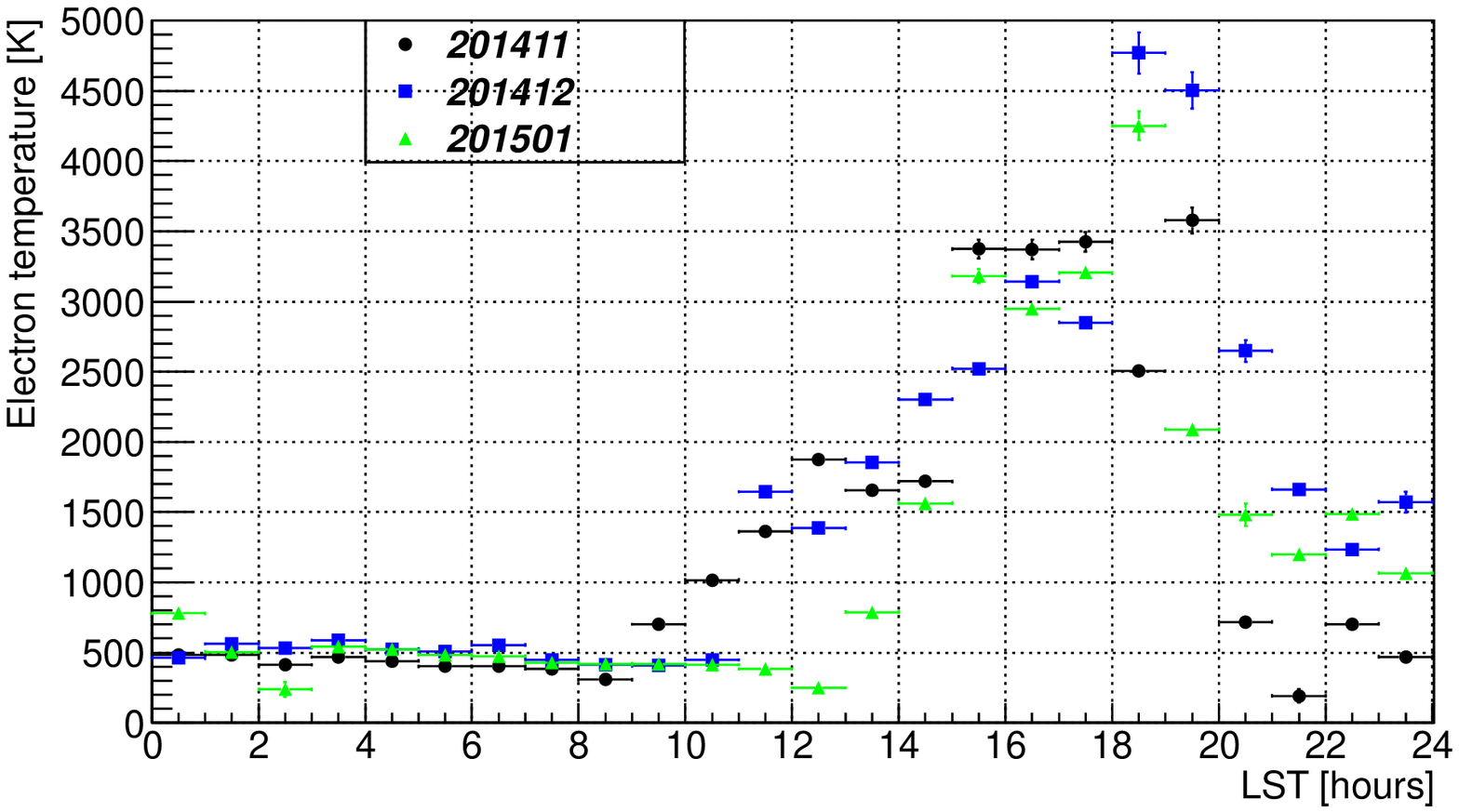}
    \caption{Electron temperature as a function of LST for Oct/Nov 2014, Dec 2014, and Jan 2015 (see Table~\ref{tab_sun} for translation between LST and local time).}
    \label{fig_te_vs_lst}
  \end{center}
\end{figure}

\red{The good correlation between the fitted parameters A and B} indicates that ionospheric absorption and emission dominate variations of the sky-averaged spectra whilst ionospheric refraction is less significant, which agrees with our initial estimates in Section~\ref{subsec_comparison_of_iono_contribs}.
As mentioned above, a very good linear correlation was typically obtained for nighttime data, but daytime data show much larger scatter (Fig.~\ref{fig_a_vs_b}~d,e,f,g).
There may be multiple reasons for this, such as non-perfect excision of data affected by the Sun or that other effects (i.e. ionospheric refraction) play a more significant role during the daytime. 
Nevertheless, we performed the same analysis for all 1\,h LST bins for three calendar months and derived $T_e$ as a function of LST (Fig.~\ref{fig_te_vs_lst}). The dependence of $T_e$ on local time will be presented in Section~\ref{sec_data_modelling}.

%

\subsubsection{Derivation of variations in ionospheric absorption}
\label{subsubsec_tau_vs_time}

\begin{figure}     
  \begin{center}   
   \includegraphics[width=3in]{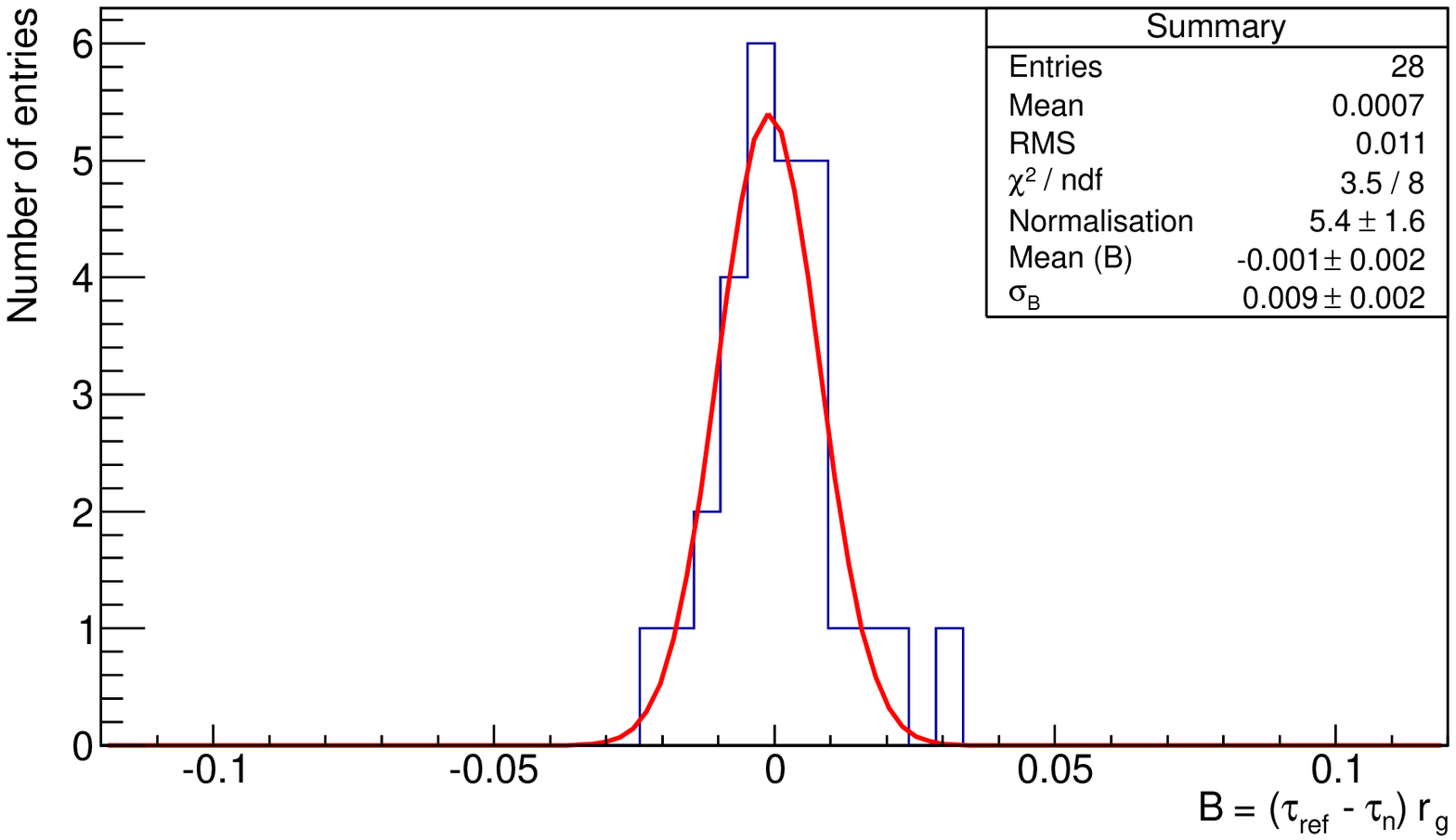}
   \caption{\red{Example distribution of fitted values of parameter $B = (\tau_{100 MHz}^r - \tau_{100 MHz}^n) r_{g}$ for the LST bin $0-1$\,hours from Oct/Nov 2014 data (X-axis values from Fig.~\ref{fig_a_vs_b}a).}} 
   \label{fig_lst0_a_histo}
  \end{center}
\end{figure}

According to equation~\ref{eq_fitted_ab_interpretation} the parameter B, fitted to differences of 1\,h LST spectra, can be interpreted as $B = (\tau_{100 MHz}^r - \tau_{100 MHz}^n) r_{g}$. Thus, its distribution represents variations in the ionospheric \red{optical depth} at 100\,MHz ($\tau_{100 MHz}$).
For every 1\,h LST bin we created a histogram of fitted parameter B.
An example distribution of B for LST bin $0-1$\,h in Oct/Nov 2014 data (distribution of X-axis in Fig.~\ref{fig_a_vs_b}a) is shown in Figure~\ref{fig_lst0_a_histo}.
Because Figure~\ref{fig_lst0_a_histo} is a distribution of differences of \red{optical depths} at 100\,MHz $(\tau_{100 MHz}^r - \tau_{100 MHz}^n)$, its standard deviation is larger by $\sqrt{2}$ than the standard deviation of the \red{optical depth} itself $\sigma_{\tau} = \sigma(\tau_{100 MHz})$.
Therefore, in order to calculate the standard deviation of $\tau_{100 MHz}$ at zenith, the calculated standard deviation of parameter B was divided by a factor $\sqrt{2}r_{g} \approx 2$, where $r_{g} \approx 1.4$ is the value of the geometric factor calculated for the conical log spiral antenna (Sec.~\ref{subsec_contrib_iono_abs}).

\begin{figure}     
  \begin{center}   
   \includegraphics[width=3in]{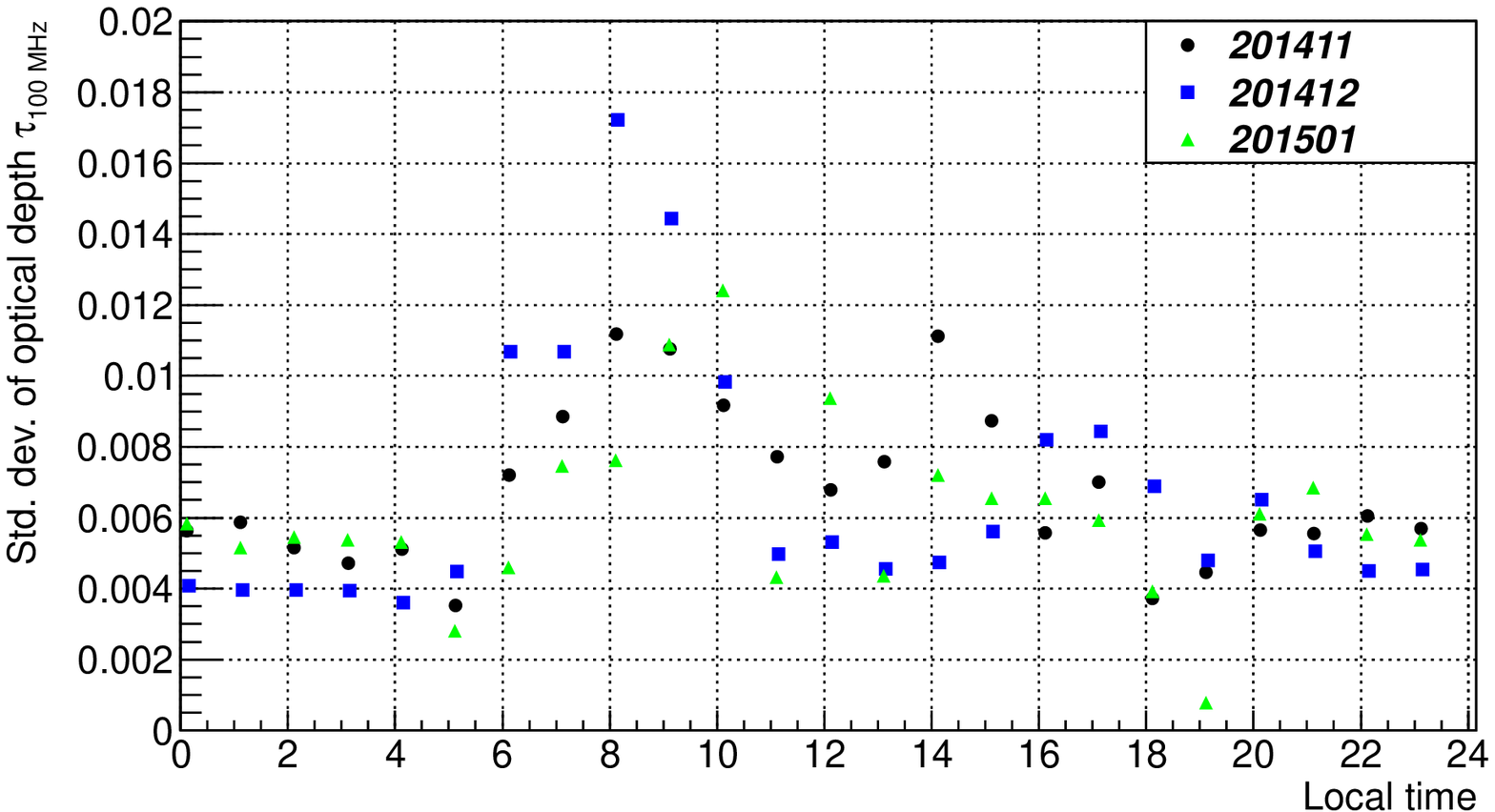}
   \caption{\red{Standard deviation of zenith \red{optical depth} at 100\,MHz as a function of local time derived for Oct/Nov 2014, Dec 2014, and Jan 2015 data.}}
   \label{fig_opacity_vs_localtime}
  \end{center}
\end{figure}

The standard deviation of \red{optical depth} $\tau_{100 MHz}$ as a function of local time calculated for all 1\,h LST bins and all four months is shown in Figure~\ref{fig_opacity_vs_localtime}. 
\red{The observed nighttime optical depth variations are typically $\sigma_{\tau} \sim$ 0.005, which corresponds to $\sim$0.002 at 150\,MHz, very similar to values measured by \citet{2014arXiv1412.2255R}. }
\red{Although our value, $\sigma_{\tau} \sim$ 0.005, is slightly larger than typically cited nighttime optical depth at zenith $\tau_{100 MHz} \sim 0.0023$ ($\sim$0.01\,dB), it is well within a range of values ($\sim0.0023 - 0.023$) cited in the literature (Sec.~\ref{subsec_contrib_iono_abs}).}
Moreover, \red{optical depth} and its variations may also be season and solar cycle dependent.

\subsubsection{Zenith absorption at 100\,MHz}
\label{subsubsec_abs_tau}

In order to measure the \red{optical depth} at zenith at 100\,MHz ($\tau_{100 MHz}$) we used a riometer-like approach. Based on multi-day observations a quiet day curve (QDC) of the maximum observed sky temperature (\red{at a state of very small residual absorption}) is determined and relative absorption is measured with respect to the QDC \red{\citep{qdc_method,belokovich}}. 

\begin{figure}
  \begin{center}
   \includegraphics[width=3in]{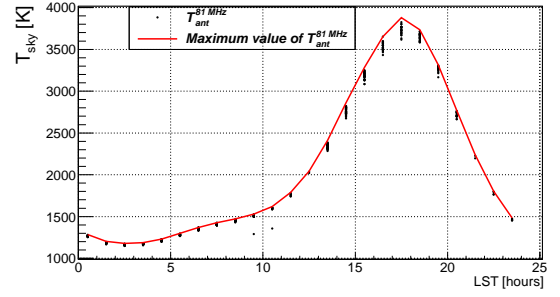}
   \caption{Quiet day curve (QDC) derived as maximum value of antenna temperature MAX($T_{ant}^{81 MHz}$), where $T_{ant}^{81 MHz}$ is a mean value of $T_{ant}$ in $80-82$\,MHz bin.}
   \label{fig_qdc}
  \end{center}
\end{figure}

Due to low-level RFI in the FM band ($88-108$\,MHz) we determined the \red{optical depth} at the RFI free frequency of 81\,MHz and scaled the result to 100\,MHz by a factor of $(81/100)^2$.
The calibrated antenna temperature at 81\,MHz ($T_{ant}^{81 MHz}$) was calculated as a mean value of $T_{ant}$ in the $80-82$\,MHz sub-band. In order to calculate absorption in the selected band, we first determined a complete (for nighttime and daytime) QDC based on 1\,h LST spectra from all the data collected between Oct 2014 and Jan 2015. 
The same quality selection criteria as described in Section~\ref{subsec_fitting_proc} were applied in order to excise data affected by any kind of variability (mainly solar or lightning).
In order to determine the QDC value for a given LST bin we used a sample of antenna temperatures $T_{ant}^{81 MHz}$ from quality-accepted 1\,h LST spectra. If for a particular LST the sample contained spectra collected during nighttime and daytime, only the nighttime bins (solar elevation $\le - 2\degree$) were used.
The QDC value $T_{ant}^{max}(LST) =  MAX(T_{ant}^{81 MHz}$) calculated for all 1\,h LST bins (including the ones collected only during daytime) is shown in Figure~\ref{fig_qdc}.
\red{We further assume that the QDC curve, $T_{ant}^{max}(LST)$, represents antenna temperature measured at a time of very small residual ionospheric absorption, and we determine relative absorption with respect to this QDC curve. }
\red{Due to limited amount of data and the fact that the state of minimum absorption occurs in the pre-dawn hours \citep{belokovich}, this assumption is valid for only some nighttime LST bins and invalid for LST bins collected during the daytime only.}

Based on equation~\ref{eq_abs_emission} we can define the difference between the maximum measured antenna temperature at 81\,MHz  $T_{ant}^{max}(LST)$ (at negligible absorption) and any other measurement $T_{ant}(LST)$ as
\begin{equation}
\begin{split}
\Delta_{qdc} = T_{ant}^{max}(LST) - T_{ant}^{n}(LST) = \\
\tau(\nu,LST) ( T_{ant}^{max}(LST) - T_{e}(LST) ),
\label{eq_qdc_diff}   
\end{split}
\end{equation}
and therefore, the \red{optical depth} at any given frequency and LST $\tau(\nu,LST)$ can be calculated from a difference between QDC and other spectra as:
\begin{equation}
\begin{split}
\tau(\nu,LST) = \frac{ T_{ant}^{max}(LST) - T_{ant}^{n}(LST) }{T_{ant}^{max}(LST) - T_{e}(LST)}, 
\label{eq_tau_from_qdc}
\end{split}   
\end{equation}
where $T_{e}(LST)$ is the electron temperature at a given LST and was calculated according to the parameterization of the data in Figure~\ref{fig_te_vs_lst}.
Equation~\ref{eq_tau_from_qdc} indicates that the accuracy of $\tau(\nu,LST)$ strongly depends on the accuracy of $T_{e}(LST)$, which is very good for the nighttime data, but varies significantly for daytime LSTs. 
Therefore, we only determined $\tau(\nu,LST)$ for LSTs which could be collected during nighttime, which excluded LST bins 14-21\,h. 

\begin{figure}
  \begin{center}
   \includegraphics[width=3in]{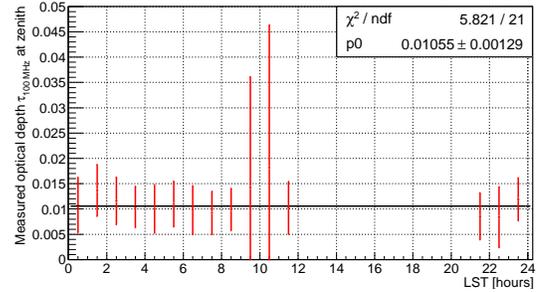}
   \caption{\red{Optical depth at 100\,MHz as a function of LST derived according to variations of antenna temperature $T_{ant}^{81 MHz}$ (mean value of $T_{ant}$ in $80-82$\,MHz bin) with respect to its maximum value $T_{ant}^{max}(LST)$. The results for 1\,h LST bins collected only during daytime were excluded from the analysis. The errors are $\approx0.005$ and therefore consistent with the optical depth variations ($\sigma_{\tau}$) derived in Section~\ref{subsubsec_tau_vs_time} using a slightly different method.}}
   \label{fig_tau_vs_lst}
  \end{center}
\end{figure}

In order to estimate a commonly cited \red{optical depth} $\tau_{100 MHz}(LST)$ at zenith we divided $\tau(81 MHz,LST)$ by the geometric factor $r^{cone}_{g}\approx$\,1.4 and subsequently multiplied by a factor $(81/100)^2 \approx 0.66$ (eq.~\ref{eq_tau_factor}). 
For each LST bin, values of $\tau_{100 MHz}$ were calculated from $T_{ant}^{81 MHz}$ derived from all 1\,h spectra and their mean and standard deviation as a function of LST are shown in Figure~\ref{fig_tau_vs_lst}.

In order to cross-check our result, we derived $\tau_{100 MHz}$ from other frequency sub-bands ($92-94$ and $102-104$\,MHz) and the results were very similar and within the measurement error.
The errors in Figure~\ref{fig_tau_vs_lst} are $\approx0.005$ and therefore consistent with the standard deviations of \red{optical depth} ($\sigma_{\tau}$) derived in Section~\ref{subsubsec_tau_vs_time} using a slightly different method.

The measured values of nighttime \red{optical depth} $\tau_{100 MHz} \sim 0.01$ (loss of 0.05\,dB) at zenith are about a factor of 4 larger \red{than the typically cited value $\sim$0.0023 ($\sim$0.01\,dB), but within a $0.0023-0.023$ ($0.01-0.1$\,dB) range of values cited in literature for nighttime \citep{evans_and_hagfors,Ionospheric_Radio,THOMPSON}.}
\red{Moreover, the measurements performed with traditional riometers did not take emission into account ($T_e(LST)=0$ in equation~\ref{eq_tau_from_qdc}). Therefore, our values of optical depth can be slightly higher than those cited in the literature, but we find the presented method more accurate, especially for measuring very small nighttime optical depths. For comparison we also calculated optical depths according to equation~\ref{eq_tau_from_qdc} with $T_e(LST)=0$ ($\tau(\nu,LST) = (T_{ant}^{max}(LST) - T_{ant}^{n}(LST) )/T_{ant}^{max}(LST)$) and it shifted the optical depths from $\sim$0.01 to $\sim$0.007.}
\red{Since nighttime absorption can also be season and solar cycle dependent, we consider our measurements to be in satisfying agreement with other measurements at the same location, but different time of year and smaller data sample \citep{2014arXiv1412.2255R}.}
Moreover, an analysis based on a larger data sample should allow us to further verify and refine our results.

\red{The presented absorption analysis can be considered as a proof of principle because it was based on a relatively small data sample (3 months worth of data), which is not sufficient for precise determination of a quiet day curve (QDC) in all LST bins. }
\red{Present analysis was strictly valid for a few nighttime LST bins. The complete QDC, built from the one year of data, will enable us to correctly determine relative absorption in all LST bins.}
\red{A complete QDC may be shifted towards higher values and result in slightly higher values of optical depth, but this should mostly affect LST bins presently observed during the daytime.} 
\red{We consider our present nighttime measurements of optical depth reliable and attribute their slightly higher than expected values primarily to time of the year and solar activity dependence.}

\section{COMPARISON WITH IONOSPHERE MODEL PREDICTIONS}
\label{sec_data_modelling}

\begin{figure}     
  \begin{center}
   \includegraphics[width=3in]{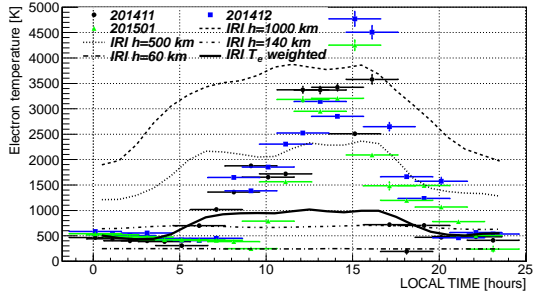}
   \caption{Comparison of measured electron temperature (data points with error bars) with predictions of International Reference Ionosphere 2012 model \citep{iri2012} weighted by \red{optical depth} (black solid line) and at four heights: 60, 140, 500 and 1000\,km (different dashed lines). \red{The bars on the time axis correspond to $\approx$3\,h local time coverage of a 1\,h LST bin over a period of 1 month (3\,h results from 1\,h LST bin itself plus $\approx$2\,h/month drift of LST with respect to local time).}}
   \label{fig_te_vs_localtime_data_with_irimodel}
  \end{center}
\end{figure}

We compared our measured electron temperatures with predictions of the International Reference Ionosphere 2012 (IRI-2012) model \citep{iri2012}. 
We initially derived the dependence of $T_e$ on LST (Fig.~\ref{fig_te_vs_lst}). Therefore, in order to derive the dependence of $T_e$ on local time for each calendar month, we converted LST to local time by applying the appropriate shift calculated for the middle of the given calendar month (Tab.~\ref{tab_sun}). 
The measured electron temperatures as a function of local time for Oct/Nov 2014, Dec 2014, and Jan 2015, over-plotted with the predictions of the IRI-2012 model for 2014-11-15 weighted by \red{optical depth} and at four different heights (60, 140, 500 and 1000\,km), are shown in Figure~\ref{fig_te_vs_localtime_data_with_irimodel}.
The overall trend of the measured and model curves is similar as they both show a rise in electron temperature at sunrise and fall at sunset, but the values of measured and predicted temperatures differ (especially during the daytime).

The weighted electron temperature ($\bar T_e$) was derived from the results of the IRI-2012 model according to the following equation:
\begin{equation}
\bar T_e = \frac{ \int_{0}^{h_{max}} T_e(h)N_e(h)\nu_c(h)dh }{ \int_{0}^{h_{max}} N_e(h)\nu_c(h)dh },
\label{eq_te_weighted}   
\end{equation}  
where $\nu_c(h)$ is a parameterization of collision rate data (Figure 7 by \citet{Aggarwal1979753}) as a function of height h, $h_{max} = 2000$\,km is the maximum height of the data generated with the IRI-2012 model, and $N_e(h)$ is electron density as a function of height obtained from the IRI-2012 model.
We found, however, that for $h \le 80$\,km the electron densities according to the IRI-2012 model were underestimated (typically $N_e=0$\,$e^{-}/m^{3}$ for nighttime). Therefore, nighttime $\bar T_e$ was originally biased towards higher values by contribution from higher ionospheric layers, because there was no (or insignificant) contribution from the D-layer.
Hence, whenever $N_e(h)$ was smaller than typical nighttime residual electron density in the D-layer, $N_0 = 10^8$\,$e^{-}/m^{3}$ \citep{evans_and_hagfors,Ionospheric_Radio,THOMPSON}, we artificially set $N_e$ to $N_0$.
This correction enabled a very good agreement between our nighttime measurements of electron temperature and weighted electron temperature derived from the IRI-2012 model. 

The measured nighttime electron temperature is approximately constant ($\approx470$\,K) and lies between model curves generated for heights $60 - 140$\,km (D and E-layers). This is the region expected to primarily contribute to the nighttime ionospheric absorption, whilst higher layers are not expected to contribute significantly to absorption during nighttime, and therefore to our measured value of the electron temperature. 
We consider our nighttime measurements as the most reliable because they are not affected by solar activity and also our results show that the assumption about the dominant role of absorption holds best for the nighttime data (Sec.~\ref{subsubsec_te_fits}). The nighttime data are also particularly interesting from the perspective of the global EoR experiments.

The electron temperatures measured during the daytime, reaching as high as $T_e\sim$3000\,K, are significantly larger than the weighted electron temperatures derived from the IRI-2012 model, which are $\bar T_e \sim$1000\,K.
The disagreement may be a combination of several factors. Firstly, our method may not be applicable to daytime data because of solar contributions affecting our data analysis (solar activity that was not excised, and movement of the Sun within the beam pattern). 
\red{Secondly, the larger contribution of F-layer refraction is not taken into account in the presented analysis.}
Thirdly, daytime measurements lie mostly between model predictions for heights $500-1000$\,km (Fig.~\ref{fig_te_vs_localtime_data_with_irimodel}), which might be a result of significant daytime absorption in the F-layer.
Thus, lack of exact knowledge of collision rates at heights h$\ge$500\,km (approximated in our calculation of $\bar T_e$ by 100\,Hz ) may lead to an underestimation of $\bar T_e$.

Therefore, we conclude that our measurements agree with the IRI-2012 model predictions (particularly at nighttime) and other measurements \citep{zhang_holt_zalucha,Pawsey} within the applicability of our method. The agreement can be tested better in the future when changing quiet-Sun conditions, un-excised solar activity remaining in the data sample, and ionospheric refraction are better accounted for in the data analysis.


\section{STABILITY TEST OF SKY DATA}
\label{sec_stddev_test}

\begin{figure}
  \begin{center}
   \includegraphics[width=3in]{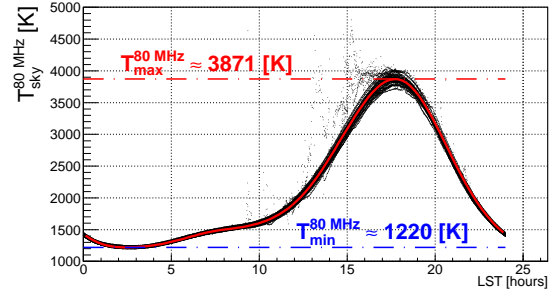}
    \caption{\red{Calibrated sky temperature $T_{ant}^{80 MHz}$ at 80\,MHz as a function of LST overplotted with its parameterization by a Fourier series (eq.~\ref{eq_fourier_series}) and the maximum ($T_{max}^{80 MHz}$) and minimum ($T_{min}^{80 MHz}$) values of the fitted Fourier series. We note that $1.9 - 3.9$\,h LST appears to be the best range in which to integrate data down for the global EoR experiment.}}
   \label{fig_tsky_vs_lst}
  \end{center}
\end{figure}

According to the dependency of $T_{ant}^{80 MHz}$ with local time (Fig.~\ref{fig_tsky_vs_lst}), we identified that $T_{ant}^{80 MHz}$ was minimal at around LST = 2.9\,hours and not at $\approx$5.8\,h when the Galactic Center was at its lowest point below the horizon.
Therefore, we decided that the optimal conditions to select spectra for averaging are at $LST_{opt}$\,$\pm$\,$\Delta LST_{opt} =2.9 \pm 1$\,hour and only during the nighttime, which results in $\lesssim$2\,h of data per day to be used for the global EoR analysis (as some data might be excised by RFI or variability criteria).

\begin{figure}
  \begin{center}
	\includegraphics[width=3in]{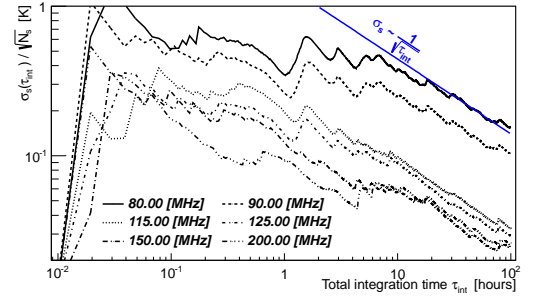}
   \caption{The standard deviation of $T_{ant}$ ($\sigma_s$) divided by the square root of the number of samples ($N_s = \tau_{int}/35$\,sec) as a function of integration time $\tau_{int}$ for several selected frequency channels. Because $\sigma_{s}$ converges and remains approximately constant when $N_s$ is sufficiently large, the standard error of the mean decreases as expected from equation~\ref{eq_radiometer_equation} by a factor $\sim1/\sqrt{N_s}$ (blue line). Therefore, the required precision may be reached after sufficiently long integration time. The test was based on 2014-10-24 - 2014-12-31 data. The data collected after 2015-01 could not be included because the optimal LST range (1.9 - 3.9\,h) began to enter daytime.}
   \label{fig_stddev_test}
  \end{center}
\end{figure}

In order to test whether the sky spectra will integrate down to the required precision, we calculated the standard deviation of $T_{ant}$ from samples of increasing numbers of integrations.
The standard deviation ($\sigma_{s}$) was calculated from a sample of spectra in the reduced and calibrated dynamic spectrum collected in the period 2014-10-24 to 2014-12-31 (in 35\,sec resolution). 
We could not include the data after 2015-01-01 because the optimal LST range (1.9 - 3.9\,h) began to enter daytime. 
Then, for several selected frequency channels, $\sigma_{s}$ was plotted as a function of integration time $\tau_{int}$, related to the number of sample integrations $N_s$ as $\tau_{int} = N_s \times 35$\,sec.
Such a test allowed us to verify that after a sufficiently long integration time, $\sigma_{s}$ converges and remains approximately constant with an increasing number of sample integrations.
Therefore, averaging an increasing number of spectra reduces the statistical error as expected from equation~\ref{eq_radiometer_equation} by a factor $\sim1/\sqrt{N_s}$ (Fig.~\ref{fig_stddev_test}), which is very encouraging and convinces us that ionospheric or instrumental effects should not prevent us from integrating down to the required precision.

\begin{figure}
  \begin{center}
   \includegraphics[width=3in]{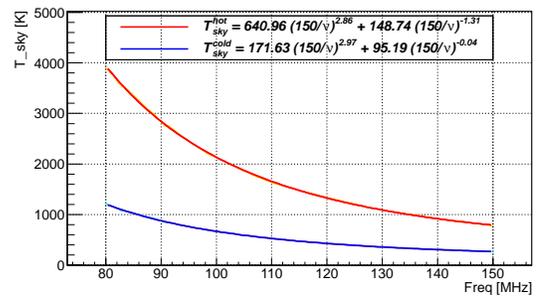}
	\caption{Sky spectra at maximum ($T_{ant}^{hot}(\nu)$) at LST$\approx$17.8\,h and minimum ($T_{ant}^{cold}(\nu)$) at LST$\sim$1.9\,h parameterized with sums of two power laws. The sum of two power laws yielded a slightly better fit than a single power law with fit residuals within $\approx$20\,K for ``hot'' and $\approx$10\,K for ``cold'' sky over most of the bandwidth.}
	\label{fig_tsky_hot_cold_fit}
  \end{center}
\end{figure}

\subsection{Expectations based on system temperature} 
\label{subsec_stddev_test_expected}

In order to verify our understanding of the $\sigma_{s}$ dependence on the integration time we generated a \red{model dynamic spectrum with separate time and frequency components and use this model to generate simulated observations}.
\red{Firstly, we parameterized the sky temperature dependence on LST ($T_{ant}(LST)$) with a Fourier series:}
\begin{equation}
\red{F(LST) = a_0 +  \sum\limits_{i=1}^6 a_i sin(2\pi i p) + \sum\limits_{i=1}^6 b_i cos(2\pi i p),}
\label{eq_fourier_series}
\end{equation}
where p is (LST/24). The data with a fitted Fourier series are shown in Figure~\ref{fig_tsky_vs_lst}\red{, where we also defined the maximum $T_{max}^{80 MHz}$ and the minimum $T_{min}^{80 MHz}$ values of the fitted Fourier series.}
\red{We define a normalized Fourier series as:}
\begin{equation}
\red{\overline{F(LST)} = \frac{F(LST) - T_{min}^{80 MHz}}{T_{max}^{80 MHz} - T_{min}^{80 MHz}}.}
\label{eq_norm_fourier_series}
\end{equation}

\red{Next we fitted a sum of two power laws in frequency to the maximum ($T_{ant}^{hot}(\nu)$, at LST$\approx$17.8\,h) and minimum ($T_{ant}^{cold}(\nu)$, at LST$\sim$1.9\,h) sky temperature (Fig.~\ref{fig_tsky_hot_cold_fit}).}
We used two power laws because the model yielded a slightly better fit than a single power law, with fit residuals within $\approx$20\,K for ``hot'' and $\approx$10\,K for ``cold'' sky over most of the bandwidth.
\red{The antenna temperature $T_{ant}^{gen}(\nu)$ was generated for the same LSTs as the sky data by adding a difference of sky spectra $(T_{ant}^{hot}(\nu) - T_{ant}^{cold}(\nu))$ modulated by $\overline{F(LST)}$ to the ``cold'' sky spectrum according to the formula:}
\begin{equation}
\begin{split}
\red{T_{ant}^{gen}(\nu) = T_{ant}^{cold}(\nu) + \overline{F(LST)} \Big(T_{ant}^{hot}(\nu) - T_{ant}^{cold}(\nu) \Big)}.
\label{eq_tsky_vs_lst_gen}
\end{split}   
\end{equation}

\red{This model represents the ``real'' sky within $\pm$20\,K, which is sufficient to estimate the errors expected from cosmic noise as a function of increasing integration time.}
\red{The statistical error on the simulated observations was calculated according the same formula which we use to calculate the error on calibrated antenna temperature derived from the reference source (equation 13 in S15, where the ratio of antenna to reference data was approximated by one and a very small contribution from the error on ambient temperature measurement was neglected):}
\begin{equation}
\begin{split}
\red{\delta T_{ant}^{gen}(\nu) = \frac{T_{ant}^{gen}(\nu) + T_{rcv}(\nu)}{\sqrt{B}}\sqrt{\frac{1}{\tau_{ant}} + \frac{1}{\tau_{ref}} },}
\label{eq_delta_tsky_vs_lst_gen}
\end{split}   
\end{equation}
\red{where $B$ is the frequency bin of $\approx$117.2\,kHz, $T_{rcv} \approx$180\,K is the receiver noise, $\tau_{ant}$=35\,sec is the integration time on the antenna, and $\tau_{ref}$=12.5\,sec is the average integration time on the reference (both from the reduced dynamic spectrum).}

\begin{figure}  
  \begin{center}
   \includegraphics[width=3in]{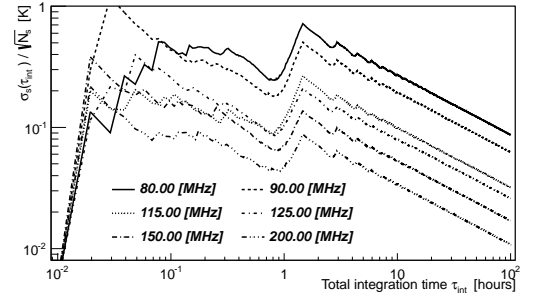}
   \caption{Standard error of the mean simulated antenna temperature ($\sigma_s/\sqrt{N_s}$) as a function of integration time $\tau_{int}$ (number of integrations in the sample can be calculated as $N_s = \tau_{int}/35$\,sec)  for the same set of frequency channels as for the same test applied to the real data (Fig.~\ref{fig_stddev_test}). The simulated data are slightly below corresponding curves in Figure~\ref{fig_stddev_test}, which might be attributed to variations introduced by the ionosphere (Fig.~\ref{fig_rms_vs_lst}) not taken into account in the simplified simulation procedure. The ripples on timescale of $\approx$2\,h are due to small changes in sky temperature in the optimal LST range (1.9 - 3.9\,h).}
   \label{fig_stddev_test_generated}
  \end{center}
\end{figure}

\begin{figure}  
  \begin{center}
	\includegraphics[width=3in]{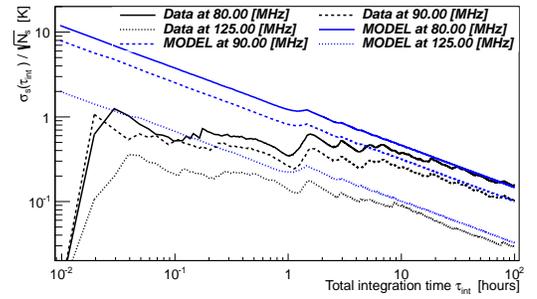}
   \caption{Standard error of the mean antenna temperature ($\sigma_s/\sqrt{N_s}$) as a function of integration time $\tau_{int}$ for sky data (black curve is the same as in Fig.~\ref{fig_stddev_test}) compared with simulated data with additional ionospheric variability (blue curve) calculated as $\sigma_{tot}/\sqrt{N_s} = (((\sigma_{s}^{gen})^2 + \sigma^2_{iono})/N_s)^{1/2}$, where $\sigma_{iono}$ was taken from Figure~\ref{fig_rms_vs_lst} at the closest frequency to the presented frequencies (80, 90 and 125\,MHz).}
   \label{fig_stddev_test_real_and_gen}
  \end{center}
\end{figure}

The dependence of the standard deviation $\sigma_{s}^{gen}$ on integration time resulting from the same procedure as previously applied to our data is shown in Figure~\ref{fig_stddev_test_generated}.
The values of $\sigma_{s}^{gen}$ converge to slightly lower values than corresponding curves in Figure~\ref{fig_stddev_test}, which might be attributed to variations introduced by the ionosphere (Fig.~\ref{fig_rms_vs_lst}) not taken into account in this simplified simulation procedure.
In order to verify this hypothesis we calculated $\sigma_{tot} = ((\sigma_{s}^{gen})^2 + \sigma^2_{iono})^{1/2}$, where $\sigma_{iono}$ was taken from Figure~\ref{fig_rms_vs_lst} at the closest frequency to those shown in Figures~\ref{fig_stddev_test}~and~\ref{fig_stddev_test_generated}.
Comparisons of $\sigma_{s}/\sqrt{N_s}$ and $\sigma_{tot}/\sqrt{N_s}$ are shown in Figure~\ref{fig_stddev_test_real_and_gen}. Although $\sigma_{tot}/\sqrt{N_s}$ is slightly above $\sigma_{s}/\sqrt{N_s}$ the overall agreement between the two is very good. 
Assuming variations shown in Figures~\ref{fig_lst0_differences_nov_oct}~and~\ref{fig_rms_vs_lst} result mainly from ionospheric variability (at least during the nighttime), which we show in Sections~\ref{sec_data_analysis}~and~\ref{sec_data_modelling}, we conclude that we understand the magnitude of sky signal variations.
Furthermore, it turns out that these variations are dominated (especially during the nighttime) by ionospheric absorption and emission.

The standard deviation calculated from the large data sample ($\sim$98\,days) converges to a constant value, which proves that, in principle, the additional variability introduced by the ionosphere does not impede integrating down to the required precision with ground-based instruments.
\red{Therefore, because ionospheric stochastic error can be averaged to a mean ionospheric contribution, the latter can be identified and removed from the sky-averaged spectrum (on a daily or hourly basis) in order to minimize the extra variance which it introduces. }
One possibility, for instance, is to identify the ionospheric contribution to the sky-averaged signal on an hourly basis (as in Fig.~\ref{fig_lst0_differences_nov_oct}). Because daily contributions to the total integration time are going to be of the order of a few hours (here we tested the LST range 1.9 - 3.9\,h), the ionospheric contribution can be identified by fitting equation~\ref{eq_diff_total} to parameterize the differences (as in Fig.~\ref{fig_ab_fit_results}) between 1\,h LST spectra and a median spectrum of this 1\,h LST bin.
Then, such a parameterization would be subtracted from the original 1\,h spectrum before averaging it together with spectra collected on other days. We are planning to identify and study the optimal procedure to subtract the ionospheric effects and systematic errors introduced by such a procedure.

\subsection{Power spectrum of ionospheric fluctuations} 
\label{subsec_power_spectrum_of_noise}

\red{Based on TEC measurements, \citet{datta_et_al} conclude that the precision required for the detection of the global EoR ($\sim$mK) cannot be achieved by a ground-based instrument because of the flicker noise nature of the ionospheric fluctuations.}
\red{The power of flicker noise is proportional to $1/f^{\alpha}$, where the exponents $\alpha$ observed in the real physical processes are in the range $0 \le \alpha \le 2$ \citep{press,milotti,johnson,PhysRevE.71.069902}}. Therefore, the power at low frequencies can be very high, but not infinite as the mathematical formula suggests.
Thus, the flicker noise does not have a well defined-mean value, which moves away from its initial value as time progresses \citep{press}. 

The flicker noise of ionospheric properties would translate into the flicker noise of the observed sky-averaged spectrum preventing even a perfect ground based instrument from integrating down to a mK precision.
However, it is expected that for real physical processes the flicker noise power law spectrum has a turn over point at some low frequency ($f_b$), breaking down to white noise, and therefore the integral of the power does not diverge at the lowest frequencies.
Many physical processes do exhibit such a break in their spectra \citep{milotti}. In such a case averaging data over long periods, much longer than the corresponding period of the turn over point ($1/f_b$), still leads to reduction in the standard error of the mean.

\begin{figure}  
  \begin{center}
	\includegraphics[width=3in]{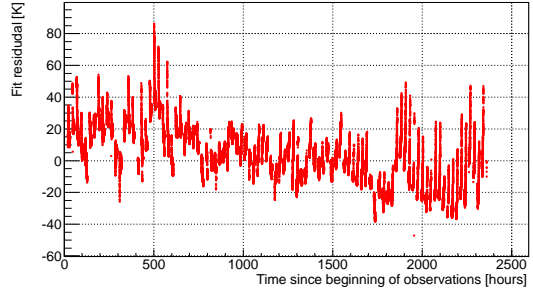}
   \caption{Nighttime fluctuations of sky temperature averaged in the $80-85$\,MHz frequency band (residuals of Fourier series fit to the observed time series).}
   \label{fig_fourier_resid}
  \end{center}
\end{figure}

Therefore, we decided to investigate the spectral characteristics of variations  observed in the sky-averaged signal (as in Fig.~\ref{fig_lst0_differences_nov_oct}), which we attribute to ionospheric variability (Sec.~\ref{subsec_fitting_proc}). 
In order to extend our analysis to the lowest possible frequencies, we used the whole three months of data ($\approx$100\,days) at $\approx$35\,sec time resolution, which allowed us to sample frequencies from $\approx0.2 \times 10^{-6}$\,Hz (period $\approx$49\,days) to $\approx 0.01$\,Hz (period $\approx$70\,sec).
We only looked at nighttime data as it is the most important from the global EoR perspective, and is not contaminated by daily solar activity. The nighttime power spectrum was averaged in the frequency band $80-85$\,MHz and its LST dependence was fitted with a Fourier series (eq.~\ref{eq_fourier_series} and Fig.~\ref{fig_tsky_vs_lst}).
Then, the Fourier series parameterization was subtracted from the original time series and the residuals are shown in Figure~\ref{fig_fourier_resid}. 

\begin{figure}  
  \begin{center}
	\includegraphics[width=3in]{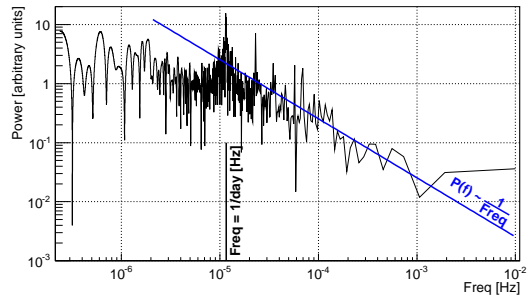}
    \caption{The power spectrum of the sky temperature fluctuations shown in Figure~\ref{fig_fourier_resid}. The frequency depenedence of the power spectrum exhibits flicker noise nature ($\sim 1/f$) of the variations at frequencies $\gtrsim 10^{-5}$\,Hz (corresponding to periods $\lesssim$1\,day). However, the power law shows the evidence of a break at lower frequencies, which makes integration to $\sim$mK precisions possible for ground-based instruments.}
   \label{fig_flicker_noise_power_spectrum}
  \end{center}
\end{figure}

The resulting time series of the fluctuations has gaps (due to daytime, RFI excision etc.). Therefore, in order to determine the power spectrum of the observed fluctuations we used the Lomb-Scargle algorithm \citep{lomb,scargle}, which is suitable for unevenly sampled data.
The power spectrum calculated by the Lomb-Scargle algorithm with the $\sim 1/f$ dependence is shown in Figure~\ref{fig_flicker_noise_power_spectrum}. 
Its frequency dependence at frequencies $\gtrsim10^{-5}$\,Hz (corresponding to period $\lesssim $1\,day) can, indeed, be very well characterized by a flicker noise.
However, the power law dependence breaks at the lowest frequencies ($\lesssim10^{-5}$\,Hz), suggesting a low frequency cut-off and therefore absence of high power contained in very long-period fluctuations in the data.
The low frequency cut-off may explain the results of our statistical test (Sec.~\ref{sec_stddev_test}), suggesting that the standard error decreases as $1/\sqrt{N_s}$ even at very long integration times, as expected in the absence of flicker noise, and therefore makes averaging of multi-day (or even multi-month) data samples to $\sim$mK precision possible.
Analysis of even larger data samples (longer time series) will enable us to verify the observed turn over point in the power spectrum at frequency $f_b \approx 10^{-5}$\,Hz.

\section{CONCLUSIONS}
\label{sec_conclusions}

Achieving the calibration precision required for the global EoR experiment is certainly a very challenging task from the instrumental and engineering points of view. 
Moreover, besides instrumental effects there are propagation effects in the ionosphere, which significantly affect the frequency structure of the foreground \citep{harish_et_al} and therefore complicate the data analysis.
The most relevant processes are ionospheric absorption, emission and refraction which, based on variability studies of ionospheric properties (such as TEC), could potentially impede detection of the global EoR from the ground \citep{datta_et_al}.
Therefore, it is important to identify the ionospheric effects in the sky-averaged radio spectrum at the frequencies of interest (50-200\,MHz) and evaluate their significance for this type of experiment.

In this paper, we presented analysis of three months of data collected between Oct 2014 and Jan 2015 with an upgraded BIGHORNS system equipped with a conical log spiral antenna and installed at the MRO.
We performed our data analysis separately on the three individual months (Oct/Nov 2014, Dec 2014 and Jan 2015).
We applied almost the same data reduction, RFI excision and calibration procedures as described in S15 and extended it with a variability criterion to excise data affected by significant variability (due to for example solar activity or lightning).

Our analysis was based on the assumption that the differences in spectra (averaged in 1\,h LST bins) collected at the same LST (particularly during the nighttime) and similar environmental conditions (i.e. ambient temperature) should represent variations due to different ionospheric conditions.
Based on the sky model integrated with the simulated antenna pattern, we estimated the expected magnitude of the ionospheric effects and compared the magnitude of absorption/emission against refraction.
The expected magnitude of these effects at frequencies above 80\,MHz is $\sim 1-100$\,K. 
\red{These estimations also led us to conclude that during the nighttime the contribution of ionospheric absorption and emission is expected to be significantly (about order of magnitude) larger than the contribution of refraction.}
The observed differences of the 1\,h LST spectra were fitted with a sum of power laws $A\nu^{-2} + B\nu^{-\alpha-2}$, which initially confirmed their ionospheric origin.

Based on the approximate linear correlation of the fitted parameters A and B, which is expected if ionospheric absorption and emission dominate over refraction, we derived the electron temperature of the ionosphere as a function of LST and local time. The measured electron temperatures agree reasonably well with predictions of the IRI-2012 model (especially during the nighttime).
Based on the QDC derived from 3 months of data, we derived the magnitude of relative absorption (\red{optical depth} $\tau \approx$0.01) and its variability ($\delta \tau \approx$0.005) as a function of local time, and these values are also in agreement with other measurements (keeping in mind they can be subject to differences due to geographical location, time of the year, or phase of the solar cycle). \red{The analysis of a larger data sample should refine these results. }
The above results lead us to conclude that we do observe ionospheric effects in our data and that their magnitude agrees with expectations. The linear correlation between the fitted parameters A and B indicates that, particularly during the nighttime, absorption and emission are indeed the dominant effects.

In order to verify whether sky-averaged spectra can be integrated down to the required precision in the presence of ionospheric effects, we have performed a statistical test on the three months of data. 
The standard deviation calculated from an increasing number of sample spectra converges to an approximately constant value after the number of sample spectra is sufficiently large. 
This result is very encouraging for ground-based global EoR experiments because it means that the statistical noise can be suppressed by a factor $1/\sqrt{N_s}$, where $N_s$ is the number of averaged spectra (or alternatively $1/\sqrt{\tau}$, where $\tau$ is the integration time).
In order to verify our understanding of the test results we applied the same test to model data (without any extra variability due to ionosphere or instrumental effects). 
The standard deviation of the model data converges to a value slightly lower than observed in the the real data. The difference corresponds to the standard deviation of intra-day variability calculated from the differences of 1\,h LST spectra, which we attributed to the ionospheric effects.
Therefore, we conclude that the observed standard deviation can be understood in terms of squared sums of standard deviations of model data (statistical noise due to the galactic foreground without any ionospheric effects) and observed intra-day variations (attributed to ionospheric effects).
Hence, the results of the test suggest that the spectra can be averaged down to the precision required by global EoR studies even with ground-based instruments.

This result is in contradiction to results (based on TEC measurements and simulations) by \citet{datta_et_al} suggesting that flicker noise (power $\sim 1/f^{\alpha}$) introduced by ionospheric variability cannot be averaged down to the required $\sim$mK precision.
Therefore, in order to understand the apparent discrepancy between these results, we determined the power spectrum of the nighttime fluctuations observed in our data.
The power spectrum indeed exhibits flicker noise dependence at frequencies $\gtrsim 10^{-5}$\,Hz, but seems to have a turn over point (not uncommon in other physical processes) at a frequency $\approx10^{-5}$\,Hz (period $\approx$1\,day) and seems to be constant (white noise) at lower frequencies.
Hence, the power present in the lowest frequencies is finite and therefore long timescale fluctuations should not impede long integrations to suppress the standard error of the mean.
We conclude that both results of the statistical (standard deviation) test and power spectrum of the observed variations suggest that the \red{ionospheric stochastic error} can be suppressed by long (multi-day or even multi-month) integrations.
\red{Therefore, the ionospheric absorption, emission and refraction and their variations are not fundamental impediments to integrate down the cosmic noise to the required $\sim$mK precision.}

Nevertheless, because the observed ionospheric effects are (as expected) 1-3 orders of magnitude larger than the global EoR signal, they have to be \red{taken into account in the data analysis} if $\sim$mK precision is to be achieved by a ground-based instrument in a reasonable time period.
In the near future, we are planning to develop an appropriate procedure and study any systematic errors introduced.




\section{ACKNOWLEDGMENTS}
This research was conducted by the Australian Research Council Centre of Excellence for All-sky Astrophysics (CAASTRO), through project number CE110001020.
This scientific work makes use of the Murchison Radio-astronomy Observatory, operated by CSIRO. We acknowledge the Wajarri Yamatji people as the traditional owners of the Observatory site.
The International Centre for Radio Astronomy Research (ICRAR) is a Joint Venture between Curtin University and the University of Western Australia, funded by the State Government of Western Australia and the Joint Venture partners.
Some of the results in this paper have been derived using the HEALPix \citep{2005ApJ...622..759G}. We would like to thank Andr\'e Offringa for customizing AOflagger software to work with BIGHORNS data. \red{We would also like to thank Matthew Francis for generating and sharing slant TEC data for the MRO.}


\bibliographystyle{apj}
\bibliography{refs}

\end{document}